\documentclass[10pt]{article}
\usepackage{fullpage,citesort,epsfig,psfrag,graphics,amsbsy,amssymb}
\usepackage{caption}

\newcommand{\bea}{\begin{eqnarray}}
\newcommand{\eea}{\end{eqnarray}}
\newcommand{\beq}{\begin{equation}}
\newcommand{\eeq}{\end{equation}}
\newcommand{\VEV}[1]{\langle#1\rangle}
\newcommand{\bfs}{\boldsymbol}

\newcommand{\be}{\begin{equation}}
\newcommand{\ee}{\end{equation}}
\newcommand{\bq}{\begin{eqnarray}}
\newcommand{\eq}{\end{eqnarray}}

\def\math{\mathsurround=0pt }
\def\leftrightarrowfill{$\math \mathord\leftarrow \mkern-6mu 
 \cleaders\hbox{$\mkern-2mu \mathord- \mkern-2mu$}\hfill
 \mkern-6mu \mathord\rightarrow$}
\def\overleftrightarrow#1{\vbox{\ialign{##\crcr
     \leftrightarrowfill\crcr\noalign{\kern-1pt\nointerlineskip}
     $\hfil\displaystyle{#1}\hfil$\crcr}}}

\def\m@th{\mathsurround=0pt }
\def\leftrightarrowfill{$\m@th \mathord\leftarrow \mkern-6mu
 \cleaders\hbox{$\mkern-2mu \mathord- \mkern-2mu$}\hfill
 \mkern-6mu \mathord\rightarrow$}
\def\overleftrightarrow#1{\vbox{\ialign{##\crcr
     \leftrightarrowfill\crcr\noalign{\kern-1pt\nointerlineskip}
     $\hfil\displaystyle{#1}\hfil$\crcr}}}

\begin{document}
\setlength{\captionmargin}{36pt}
\begin{titlepage}
\begin{flushright}
UFIFT-HEP-07-10\\
\end{flushright}

\vskip 3cm

\begin{center}
\begin{Large}
{\bf 1- Brane Sources for the Lightcone Worldsheet:\\
 $Q$-branion  - ${\bar Q}$-branion  
Scattering to One Loop
}
\end{Large}

\vskip 2cm
{\large 
Charles B. Thorn\footnote{E-mail  address: {\tt thorn@phys.ufl.edu}}
}
\vskip0.20cm
{\it Institute for Fundamental Theory\\
Department of Physics, University of Florida,
Gainesville FL 32611}

\vskip 1.0cm
\end{center}

\begin{abstract}
\noindent This paper extends the study, initiated by Rozowsky
and Thorn \cite{rozowskyt}, of gauge fields in interaction with
Dirac fields living on separated parallel 1-branes. In a lightcone
description, replacing static point sources by 1-brane sources 
allows $p^+$ conservation to be maintained in their presence,
which simplifies the lightcone quantization procedure. 
Here we calculate on-shell
branion scattering amplitudes through 1 loop in lightcone gauge, and thereby
resolve a puzzling ambiguity encountered in the off-shell
calculations of \cite{rozowskyt}. We confirm that 
infrared divergences cancel in
properly defined scattering probabilities. This work 
lays the groundwork for the incorporation of 1-brane sources in the
lightcone worldsheet formalism.
\end{abstract}
\vfill
\end{titlepage}
\section{Introduction}
The response of gauge fields to 
separated static quark antiquark ($Q{\bar Q}$) sources provides valuable
information about a gauge theory. For example, in QCD this response
is described by the expectation of a long rectangular Wilson loop 
$\VEV{W(L,T)}$, with $T\gg L$, which provides among other things
an elegant criterion for quark confinement: In pure Yang-Mills
without quarks, $\VEV{W(L,T)}\sim
e^{-T_0LT}\left[1+O(\sum_n c_ne^{-\Delta_n T})\right]$ 
implies a constant confining force $T_0$. With quarks included,
this criterion only works in the 't Hooft limit \cite{thooftlargen}
$N_c\to\infty$ of QCD generalized
from a 3 color to an $N_c$ color gauge theory.\footnote{At finite 
$N_c$ the production of a
quark antiquark pair will break the flux tube and the
area law would fail.}  
The confining force is due to the formation of a flux tube, whose
excitations can be further studied by extracting the discrete 
(at $N_c=\infty$ \cite{klebanovmt}) excited energy
levels $\Delta_n$ from an analysis of the large $T$ behavior of $\VEV{W}$.

Such a system of sources is also an insightful 
tool in the study of String/Field
duality as exemplified by the AdS/CFT correspondence \cite{maldacena}.
On the field theory side the CFT is the conformally invariant 
${\cal N}=4$ super Yang-Mills
theory. Its response to the $Q{\bar Q}$ source system can be
calculated at weak 't Hooft coupling $\lambda\equiv{N_c\alpha_s/\pi}\ll1$
by expanding $\VEV{W}$ as a sum of planar Feynman diagrams. 
On the string side, $\VEV{W}$ is given as a worldsheet path
integral for an open string, moving on a manifold AdS$_5\times$S$^5$,
and whose ends are fixed to two points separated by a distance $L$ on the
boundary of $AdS_5$. At strong 't Hooft coupling the worldsheet
dynamics can be treated semi-classically, enabling the calculation of 
the ground state energy $-c\sqrt{\lambda}/L$ of the flux 
tube \cite{maldacenaqqbar}
as well as its excited energy spectrum \cite{callang,klebanovmt}
when $\lambda\gg1$.

Although the AdS/CFT correspondence asserts the equivalence of ${\cal N}=4$
Yang-Mills to IIB superstring theory on AdS$_5\times$S$^5$ at
all couplings, very little is known about the physics, 
from either point of view, at intermediate coupling $\lambda=O(1)$. 
Some physical quantities (e.g. BPS states) are protected
by supersymmetry from dependence on the coupling, and so are ``known''
at all coupling. An example is the single straight Wilson line representing
a static isolated quark. The circular Wilson loop, which is conformally
related to the Wilson line and ``almost'' BPS, does
depend on the coupling, but because all diagrams except rainbow
graphs cancel it is easily computable
by graph summation at all coupling \cite{grossd,ericksonsz}. 
Unfortunately, the rectangular
Wilson loop does not enjoy these cancellations. Nonetheless, some
interesting qualitative insight into its behavior has been obtained by 
summing the ladder subset of planar diagrams 
\cite{ericksonssz,browertt,klebanovmt}.

In a separate line of development, the lightcone worldsheet formalism
\cite{bardakcit,thorngauge,gudmundssontt} has provided
a way to map, in a generic way, the sum of all planar diagrams of a 
wide range of quantum field
theories to a worldsheet dynamical system. Treated in mean field theory,
a plausible approximation in the strong 't Hooft coupling limit, 
this worldsheet system resembles a string moving on an AdS-like
manifold \cite{bardakcitmean,bardakcimean,trant}, encouraging the hope that
a more exact treatment of it can help in understanding String/Field
duality at all coupling. 
Our aim in the present article is to take a first step toward including 
$Q{\bar Q}$ sources in this formalism, a task that involves 
complications which we briefly describe below. 

The first complication is the awkwardness of describing a fixed 
point source on the lightcone. Lightcone time is $\tau=(t+z)/\sqrt2$.
This leaves ${\bfs x}=(x, y)$ and $x^-=(t-z)/\sqrt2$ as spatial coordinates.
We would like a point source to be at fixed $x,y,z$ not fixed $x,y,x^-$.
Fixed $x^-$ would describe an object moving at the speed of light,
and a point source at fixed $(x,y,\tau-x^-)$
would not be static with respect to lightcone time. 
Furthermore, either alternative would
violate $p^+$ conservation,
an essential ingredient of the natural lightcone symmetry of the bulk
gauge theory, Galilei invariance in the transverse space. 
Since the momentum component $p^+=(E+p^z)/\sqrt2$ plays the role of the 
Newtonian mass in Galilei boosts, Galilei invariance dictates
its conservation.
A way to maintain $p^+$ conservation was proposed
in \cite{rozowskyt}: fix the transverse
coordinate of the source, but allow the source to move freely on a line
parallel to the $z$-axis. In the language of string theory the
source is then not a point (0-brane) but a 1-brane. For brevity
we have called particles 
living on  a 1-brane branions. So in \cite{rozowskyt} we replaced the 
usual static $Q{\bar Q}$
system with a branion in color irrep $N_c$ living on a 1-brane together with
a branion in color irrep ${\bar N}_c$ living on a second  
1-brane separated from the first by a distance $L$. Then $p^+$ 
conservation will be preserved at the price of having to solve the
limited 1+1 dynamics of the branions. This generalization of sources
is also natural for the lightcone parametrization of string which
relies on $p^+$ to label points on a string. In this parametrization
the motion of the string in $x^-(\sigma,\tau)$ 
is completely constrained in terms of the transverse motion
${\bfs x}(\sigma,\tau)$. Fixing ${\bfs x}$ at the string ends allows
no freedom to independently fix boundary conditions on $x^-$. 
Indeed $x^-$ will have Neumann
boundary conditions whether ${\bfs x}$ has Dirichlet or Neumann
boundary conditions. 

The second complication, present even using 1-brane sources, is that the 
lightcone worldsheet formalism is constructed from the planar
diagrams using transverse momentum space in the Feynman rules.
As a consequence the worldsheet path integral is expressed in
terms of ${\bfs q}(\sigma,\tau)$ which is T-dual to the usual
transverse coordinates ${\bfs x}(\sigma,\tau)$: 
${\bfs q}^\prime = \dot{\bfs x}$.
(The corresponding relation between $\dot{\bfs q}$ and ${\bfs x}^\prime$ is
complicated, depending on the detailed dynamics.)
This is fundamental to the formalism, which is founded on
the worldsheet representation of a gluon propagator:
\bea
\exp\left\{-i{x^+\over2p^+}
{{\boldsymbol{p}}^2}\right\}&=&
\int_{{{\boldsymbol{q}}(0,\tau)=0\atop
{{\boldsymbol{q}}(p^+,\tau)={\boldsymbol{p}}}}} 
DcDbD{\boldsymbol{q}}\ \exp\left\{\int_0^T d\tau\int_0^{p^+} d\sigma
\left(b^\prime c^\prime -{1\over2}
{\boldsymbol q}^{\prime2}\right)\right\}
\label{masterform}
\eea
Here $\sigma,\tau$ parameter space is a rectangle
$0\leq \tau\leq T\equiv ix^+$, $0\leq \sigma\leq p^+$
and ${\bfs q}(\sigma,\tau)$ is a worldsheet field satisfying
Dirichlet boundary conditions such that 
${\bfs q}(p^+,\tau)-{\bfs q}(0,\tau)={\bfs p}$. The
Grassmann $b,c$ ghost
path integral cancels the determinant prefactor coming from the
Gaussian ${\bfs q}$ path integration. Replacing all the propagators
in a planar Feynman diagram with this representation automatically constructs
the diagram's worldsheet representation. The complication with introducing
localized 1-brane sources at the worldsheet boundaries 
is that, in the string representation, the fixed 
1-brane locations are ${\bfs x}(0,\tau),{\bfs x}(p^+,\tau)$. The fact that
they are static locations translates to simple Neumann conditions 
on the dual variables ${\bfs q}^\prime=0$. This is not so bad. The
complication comes in describing the separation between the branes
\bea
{\bfs L}= {\bfs x}(p^+,\tau) -{\bfs x}(0,\tau)
=\int_0^{p^+}d\sigma {\bfs x}^\prime
\eea
For a string in flat space ${\bfs x}^\prime = -\dot{\bfs q}$ and
the separation can be interpreted as a non-zero ``momentum''
associated with translational invariance in ${\bfs q}$.
But the worldsheet action derived from the sum of planar diagrams
shows that ${\bfs q}$ has, in general, very complicated interactions with
other worldsheet degrees of freedom that are not necessarily 
interpreted as coordinates of a manifold, as they happily can be
in the AdS/CFT case.

Although we will leave definitive resolution of these difficulties to 
future work, we can catch a glimpse of the issues involved by
considering the lightcone description of an AdS string
\cite{metsaevtt}. We choose coordinates so that the line element in AdS
is $ds^2={R^2}(dx_\mu dx^\mu+dz^2)/z^2$. Then the worldsheet action
for a string moving on AdS$_5$ is
\bea S_{ws}\equiv\int
d^2\xi{\cal L}=-{T_0\over2}\int d^2\xi\sqrt{g}g^{\alpha\beta}
{R^2\over z^2}(\partial_\alpha x\cdot\partial_\beta x+\partial_\alpha z
\partial_\beta z)\nonumber
\eea
For ${\cal N}=4$ super Yang-Mills, 
$T_0R^2=\sqrt{g^2N_c/4\pi^2}=\sqrt{\lambda}$.
Lightcone parametrization of the string means $x^+=\tau$ and 
${\cal P}^+=1$, where ${\cal P}^+$ is the momentum conjugate to $x^-$.
Then in this parametrization
\bea
S_{ws}&\to& \int d\tau \int_0^{p^+} d\sigma 
{1\over2}\left[{\dot{\boldsymbol{x}}}^2+\dot{z}^2-{R^4T_0^2\over z^4}(
{\boldsymbol x}^{\prime2}+z^{\prime2})
\right]\nonumber
\eea
For a closed string one must also impose the constraint
$\int_0^{p^+} d\sigma ({\bfs x}^\prime\cdot{\bfs{\cal P}}+z^\prime \Pi)=0$.
The equation of motion for ${\bfs x}$ following from this action is 
\bea
{\ddot{\bfs x}}&=&
\left({R^4T_0^2{\bfs x}^\prime\over z^4}\right)^\prime\nonumber
\eea
To put the AdS string action in a form similar to the lightcone
worldsheet action read off from graph summation, 
we do the T-dual transformation
\bea
{\bfs q}^\prime=\dot{\bfs x},\qquad \dot{\bfs q}={R^4T_0^2\over z^4}
{\bfs x}^\prime\nonumber
\eea
The integrability condition for these equations implies
the equation of motion for ${\bfs x}$. Expressing the worldsheet Lagrangian in
terms of ${\bfs q}$ gives\footnote{The Lagrangian for T-dual variables
is not obtained by direct substitution into the
Lagrangian for the original variables, 
as was erroneously done in v1 of this eprint. That error and several 
typos are corrected in the followint equation.
T-duality for the action is best
understood using the phase space action principle \cite{goddardgrt,metsaevtt}.
The canonical momentum to $(x^\mu, z)$ is $({\cal P}^\mu,\Pi)
=-R^2T_0\sqrt{g}g^{0\beta}\partial_\beta (x^\mu,z)/z^2$. 
%Then the 
%canonical hamiltonian density is
%\bea
%{\cal H}
%&=&{\kappa\over2}\left({z^2\over T_0R^2}({\cal P}^2+\Pi^2)
%+{T_0R^2\over z^2}(x^{\prime2}+z^{\prime2})\right)
%+\mu(x^\prime\cdot{\cal P}+z^\prime\Pi)
%\eea
Then the phase space Lagrange density is
\bea
{\cal L}
&=&{\dot x}\cdot{\cal P}+{\dot z}\Pi
-{\kappa\over2}\left({z^2\over T_0R^2}({\cal P}^2+\Pi^2)
+{T_0R^2\over z^2}(x^{\prime2}+z^{\prime2})\right)-\mu(x^\prime\cdot{\cal P}+z^\prime\Pi)
\eea
where $\kappa=-(\sqrt{g}g^{00})^{-1}$, $\mu=-{g^{01}/g^{00}}$.
To discuss T-duality, simply replace ${\cal P}=q^\prime$ \cite{bardakcit}.
\bea
{\cal L}
%&=&{\dot x}\cdot q^\prime+{\dot z}\Pi
%-{\kappa\over2}\left({z^2\over T_0R^2}(q^{\prime2}+\Pi^2)
%+{T_0R^2\over z^2}(x^{\prime2}+z^{\prime2})\right)-\mu(x^\prime\cdot
%q^\prime+z^\prime\Pi)\nonumber\\
&=&{\dot x}\cdot q^\prime
-{\kappa\over2}\left({z^2\over T_0R^2}q^{\prime2}
+{T_0R^2\over z^2}x^{\prime2}\right)-\mu x^\prime\cdot
q^\prime+{T_0R^2\over2z^2}\left[
{({\dot z}-\mu z^\prime)^2\over \kappa}-\kappa z^{\prime2}\right]
\eea
where $\Pi$ has been algebraically eliminated, 
and we immediately see a symmetry up to surface terms 
under $x\leftrightarrow q/T_0$, $z\to R^2/z$. In the form written, ${\cal L}$ 
gives a good action principle for the variable $x$. To get a good action 
principle for the variable $q$, the first term should be replaced
by ${\dot q}\cdot x^\prime$. We find
\bea
{\dot x}q^\prime-{\dot q}x^\prime={\partial\over\partial\tau}
({x}\cdot q^\prime)-{\partial\over\partial\sigma}
({x}\cdot {\dot q})=-{\partial\over\partial\tau}
({q}\cdot x^\prime)+{\partial\over\partial\sigma}
({q}\cdot {\dot x})
\eea
so this can be achieved by adding a surface term to the action.}
\bea
{\cal L} \to {1\over2}\left[-{\bfs q}^{\prime2}
+{z^4\over R^4T_0^2}\dot{\bfs q}^2
+\dot{z}^2
-{R^4T_0^2\over z^4}z^{\prime2}\right]\nonumber
\eea
We recognize in the ${\bfs q}^{\prime2}$ term the part of the
QFT worldsheet action coming from the propagator representation
(\ref{masterform}). The rest of the AdS worldsheet action
must simulate the sum over planar loop corrections. Notice that the
$\dot{\bfs q}$ dependence is negligible near the boundary of AdS ($z=0$).
The intuitive
origin of such terms is explained in the foundational papers
on the lightcone worldsheet \cite{bardakcit,thorngauge,gudmundssontt}. 
We just mention here that
a loop is represented on the QFT worldsheet by a line segment at
fixed $\sigma$ on which $\dot{\bfs q}=0$. Thus terms in the action
that energetically favor this condition will be gradually brought
into the worldsheet action as one includes more and more loops.
It is very plausible that in the strong 't Hooft coupling limit a
mean field treatment of the sum over loops can be represented
by a bulk term in the action similar to the ${\dot{\bfs q}}^2$
term in the AdS string action.
Our purpose here is to give an indication of
how to describe separated 1-branes using the lightcone worldsheet formalism.
From the T-duality transform we see that a string ending on
two 1-branes separated by ${\bfs L}$ must satisfy the
constraint
\bea
{\bfs L}=\int_0^{p^+}d\sigma{z^4\over T_0^2R^4}\dot{\bfs q}
\label{separation}
\eea  
In fact this quantity is conserved by the AdS dynamics and so it
is a constraint that imposed initially will hold for
all times thereafter. However, its analogue in
the QFT worldsheet will depend in detail on the outcome of the sum over loops,
and is not expected in a generic theory to have at all the
simplicity of this formula. On the other hand, from the
point of view of Feynman diagrams in quantum field theory, 
there is no doubt about how the 1-brane separation enters.
Quantum fields, representing branions, in 1+1 space-time 
dimensions will live on 
each 1-brane, and these fields will interact with the
gauge fields in the bulk. A gluon propagator that ends
on a branion will have that end localized on the corresponding
brane. If the gluon propagator is expressed in transverse
momentum space, this means
the gluon branion vertex will be associated with a factor
$e^{i{\bfs Q}\cdot{\bfs r}}$ where ${\bfs r}$ is the transverse
location of the 1-brane and ${\bfs Q}$ is the gluon momentum. 
We expect it to be a significant
challenge, beyond the scope of this
article, to figure out in detail how this simple prescription turns into
a constraint like (\ref{separation}) in the lightcone worldsheet
formalism. 

We devote the remainder of this article to
a study of the branion-gauge field interactions
at weak coupling, i.e. to the evaluation of the
corresponding Feynman diagrams through
1 loop. Although we shall not attempt to give a definitive
interpretation of our results in terms of the lightcone
worldsheet here, we shall take a step in that direction
by regularizing the loop integrals in a
worldsheet friendly way. We employ dual momentum variables,
and the ultraviolet cutoff $e^{-\delta{\bfs q}^2}$ as in
\cite{bardakcitmean,bardakcimean,thornscalar,chakrabartiqt1,chakrabartiqt2}. 
We shall extend
the results of \cite{rozowskyt} in important ways. In the
latter work four branion 1-loop Feynman diagrams were evaluated
in lightcone gauge, but with the branions off-shell. A simple ultraviolet
cutoff on transverse momentum was imposed, and infrared divergences
were regulated by discretizing $p^+$. This is
natural from the point of view of lightcone worldsheet path integrals,
because it is nothing more nor less than defining the path integrals on a
lattice  \cite{gilest,casher,thornfishnets} -- a very standard thing to do. 
However, in \cite{rozowskyt} we examined the
continuous $p^+$ limit for the off-shell amplitudes we computed,
and found some residual artificial $p^+=0$ divergences. These
are artificial because true infrared divergences are not present off-shell.
On the other hand off-shell amplitudes are gauge non-invariant and
unphysical, so such artificial divergences are not ruled out.
Unfortunately, in \cite{rozowskyt} it was found
that these divergences did not disappear unambiguously
in the on-shell limit: this limit involved
quantities of the form $0/0$, with values that depended on
exactly how the on-shell limit was taken. This issue
was left unresolved in \cite{rozowskyt}, 
because it was also tangled up with conventional
infrared divergences which were beyond the scope of that paper.

In the work described here we calculate on-shell
scattering amplitudes, using discretized $p^+$ as an infrared
cutoff that makes these on-shell quantities finite. Then we
do the standard Lee-Nauenberg analysis of infrared divergences
and show that they cancel as they should, allowing an
unambiguous continuum limit of the $p^+$ sums. The resolution
of the ambiguity found in \cite{rozowskyt} is that the on-shell limit
and continuous $p^+$ limit do not commute: one must only take 
$p^+$ continuous for physical on-shell quantities.
In the course of these calculations, we identify all of the 
counterterms that are needed to remove gauge 
violating artifacts that crop up because of ultraviolet divergences.
We identify these counterterms by comparing the results of
our $\delta$ regularization to the results given by  dimensional
regularization in the transverse dimensionality. The assumption
here is that dimensional regularization gives the correct
gauge invariant results. In fact, one of these inferred
counterterms is essential for the cancellation of infrared divergences,
giving some independent support for this assumption.
This last counterterm shows worldsheet non-local features when directly
interpreted. However, as in the case of some of the counterterms
needed in the gluon scattering calculations
of \cite{chakrabartiqt2} it is possible to realize them locally
if additional worldsheet fields are introduced.

The rest of the paper is organized as follows. In Section 2
we summarize the Feynman rules for branions in interaction
with bulk gauge fields and give the tree level branion scattering amplitude.
In the next three sections we evaluate the branion and gluon self energy
diagrams, triangle diagrams and box diagrams respectively.
In section 6 we show that the residual infrared divergences
in the one loop elastic scattering amplitudes cancel in their
contribution to scattering probabilities against divergences
in the probability for the emission of extra soft gluons.
Concluding remarks are in Section 7.
Finally, there are two appendices in which needed loop integrals
are evaluated.

\section{Feynman Rules for Branions and 4 Branion Trees}
\begin{figure}[ht]
\begin{center}
\begin{tabular}{|c|c|}
\hline
\multicolumn{2}{|c|}{\bf Light-Cone Feynman Rules} \\
\hline
&\\[-.25cm]
$
\begin{array}[c]{c}
\psfig{file=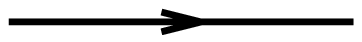,width=0.45in} 
\end{array} 
$
& $-{i\over \gamma^\alpha p_\alpha+m}$ \\[.25cm]
\hline
&\\[-.25cm]
$
\begin{array}[c]{c}
\psfig{file=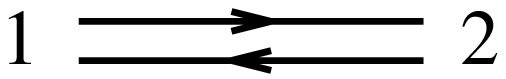,width=0.55in} 
\end{array} 
$
& $ -{i\over K^2}\left(\eta^{\mu_1\mu_2}-{K^{\mu_1}\eta^{\mu_2+}
+K^{\mu_2}\eta^{\mu_1+} \over K^+} \right)$ \\[.25cm]
\hline
&\\[-.25cm]
$
\begin{array}[c]{c}
\psfig{file=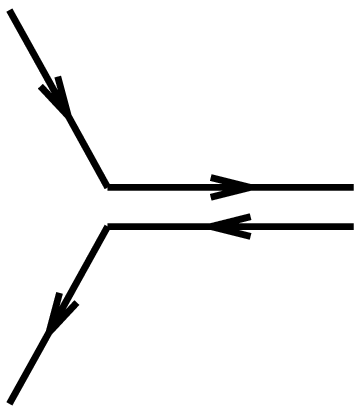,width=0.45in} 
\end{array} 
$
& $ ig\gamma^{\alpha} $ \\
\hline
&\\[-.25cm]
$
\begin{array}[c]{c}
\psfig{file=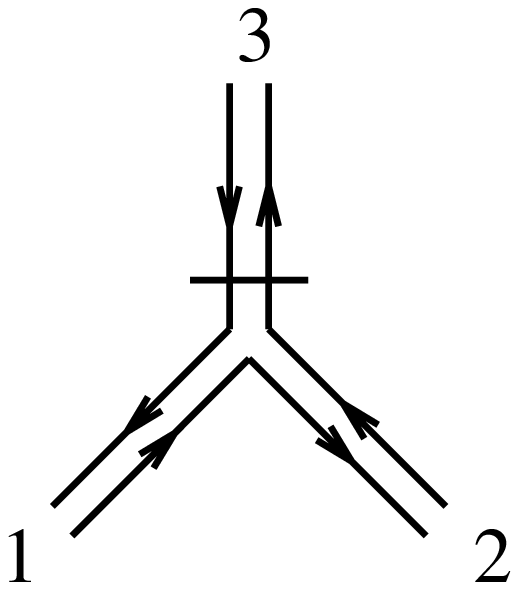,width=0.5in} 
\end{array}
$
& $ -ig\,\eta^{\mu_1 \mu_2}(Q_1-Q_2)^{\mu_3} $ \\
\hline
&\\[-.25cm]
$
\begin{array}[c]{c}
\psfig{file=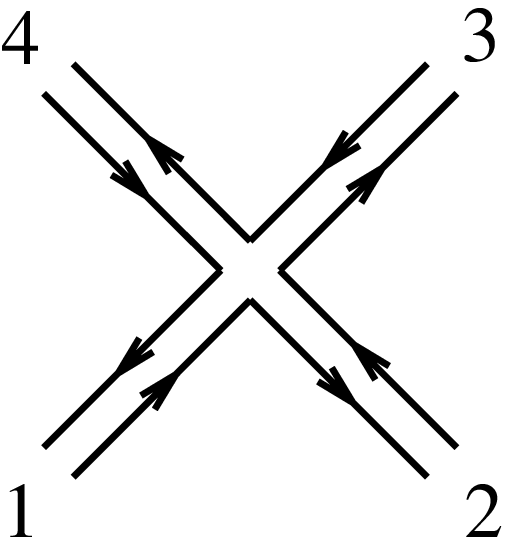,width=0.5in} 
\end{array} 
$
& $ ig^2\left[2\eta^{\mu_1 \mu_3}\eta^{\mu_2 \mu_4}
-\eta^{\mu_1 \mu_2}\eta^{\mu_3 \mu_4}
-\eta^{\mu_1 \mu_4}\eta^{\mu_2 \mu_3}\right] $ \\
\hline
\end{tabular}
\end{center}
\caption{Light-cone Feynman rules using ``double line'' notation. All
momenta in vertices are taken to be incoming and the line in the
three-gluon vertex distinguishes the three cyclic orderings. Index
$\alpha$ only includes brane coordinates, while indices $\mu_i$ run
over all coordinates. }
\label{FeynmanRules}
\end{figure}

The lightcone setup and lightcone gauge Feynman rules for
branions, taken to be 1+1 Dirac fermions, were obtained in 
\cite{rozowskyt} and summarized
in a table of that reference reproduced here in Fig.~\ref{FeynmanRules} for
the reader's convenience.
The physical process we analyze in this article is on-shell
branion-branion scattering, where the two incoming branions as
well as the two outgoing branions are on different 1-branes
separated by a distance ${\bfs L}$. We write the amplitude for this process
as a Fourier transform:
\bea
{\tilde\Gamma}(p,q,Q_\|,{\bfs L})\equiv\int{d{\bfs Q}\over (2\pi)^2}
e^{i{\bfs Q}\cdot{\bfs L}}\Gamma(p,q,Q_\|,{\bfs Q})
\eea
Here $p,q$ are the 2-vector momenta of the two incoming branions and
$Q_\|$ is the momentum transfer of the process, the final 2-vector
momenta being $p+Q_\|,q-Q_\|$ respectively. The integrand $\Gamma$
is evaluated by the usual momentum space Feynman rules, with the
understanding that the branions can absorb or give up any amount
of transverse momentum with no change of state. The gluons attached
to the right branion carry away a total transverse momentum of ${\bfs Q}$
which is absorbed by the left branion from the gluons attached to it.
We calculate $\Gamma$ by fixing this total transverse momentum and
integrating over all the other momenta as loop momenta.

With this understanding, we find for the lowest order (tree) 
contribution to this
process
\bea
\Gamma^{\rm Tree}
={2ig^2\gamma_1^+\gamma_2^+ Q^-\over Q^+Q^2}
={2ig^2\gamma_1^+\gamma_2^+ Q^-\over Q^+({\bfs Q}^2-2Q^+Q^-)}
\eea
Because the branions are free to move in only one dimension the
on-shell condition is very restrictive: there is only the option
of forward and backward scattering: $Q_\|=0,q-p$ respectively.
To avoid $Q^+=0$ issues we restrict consideration in the rest of the
paper to on-shell backward scattering, $Q^+=q^+-p^+$, 
$Q^-=-m^2{(q^+-p^+)/2q^+p^+}$. Then
\bea
\Gamma^{\rm Tree}
&=&-{ig^2m^2\gamma_1^+\gamma_2^+ \over p^+q^+{\bfs Q}^2+m^2(q^+-p^+)^2}\\
{\tilde\Gamma}^{\rm Tree}&=&-{ig^2m^2\gamma_1^+\gamma_2^+ \over 2\pi p^+q^+}
K_0\left(Lm{q^+-p^+\over\sqrt{p^+q^+}}\right)
\eea
It is easy enough to obtain ${\tilde\Gamma}$ for the tree amplitude
in terms of the Kelvin function $K_0$. 
But for the 1 loop calculations that follow we calculate $\Gamma$
and do not carry out the final Fourier transformation that would 
convert it to ${\tilde\Gamma}$.
%\newpage
\section{Self-Energy Diagrams}
\subsection{Branion Self Energy}
The branion self energy diagram is shown in 
\begin{figure}[ht]
\begin{center}
\psfrag{'p'}{$p$}
\psfrag{'p-k'}{$p-K_\|$}
\psfrag{'k'}{${K}$}
\includegraphics[width=5cm]{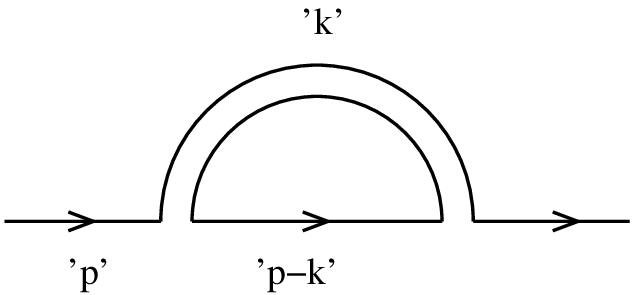}
\caption{Branion self-energy diagram. Only the gluon line carries transverse
momentum.}
\label{branionse}
\end{center}
\end{figure}
Fig.~\ref{branionse}.
Notice that the gluon propagates in the bulk whereas the fermion
resides on the 1-brane. Applying the Feynman rules Fig.~\ref{FeynmanRules}
we find
\bea
-i\Sigma(p)&=&g^2N_c\int{d^4K\over(2\pi)^4}e^{-\delta({\bfs K}+{\bfs k}_0)^2}
{\gamma^+(m-\gamma\cdot(p-K_\|))\gamma^+\over m^2+(p-K_\|)^2}
{2K^-\over K^+ K^2}\nonumber\\
&=&g^2N_c\gamma^+\int{d^4K\over(2\pi)^4}e^{-\delta({\bfs K}+{\bfs k}_0)^2}
{2(p^+-K^+)\over m^2+(p-K_\|)^2}
{2K^-\over K^+ K^2}
\eea
To evaluate the $K^-$ integral by residues we must add a semi-circle 
at infinity that gives a finite contribution, since the integrand
only falls as $1/K^-$ at large $K^-$:
\bea
{2(p^+-K^+)\over m^2+(p-K_\|)^2}{2K^-\over K^+ K^2}
\sim{2(p^+-K^+)\over 2(p^+-K^+)K^-}{2K^-\over (-2)K^{+2}K^-}
\sim-{1\over K^{+2}K^-}
\eea
Thus the added semi-circular contour will contribute $-i\pi/K^{+2}$
if it closes the contour in the upper half plane and $+i\pi/K^{+2}$
if it closes in the lower half plane. In evaluating the  $K^-$ integral by
residues it is convenient to close in the upper half plane when
$K^+>p^+$ and in the lower half plane when $K^+<p^+$. The integral over 
$K^-$ will be given by the residues of any poles inside the
closed contour minus the contributions of the added semi-circular contours:
\bea
-i\Sigma(p)&=&g^2N_c\gamma^+\int{d{\bfs K}\over(2\pi)^4}
e^{-\delta({\bfs K}+{\bfs k}_0)^2}
\Bigg[\sum_{K^+}
{\rm Residue}{2(p^+-K^+)\over m^2+(p-K_\|)^2}{2K^-\over K^+ K^2}
\nonumber\\
&&+i\pi\sum_{K^+>p^+}{1\over K^{+2}}-i\pi\sum_{K^+<p^+}{1\over K^{+2}}\Bigg]
\nonumber\\
&=&g^2N_c\gamma^+\int{d{\bfs K}\over(2\pi)^4}
e^{-\delta({\bfs K}+{\bfs k}_0)^2}
\Bigg[\sum_{K^+}
{\rm Residue}{2(p^+-K^+)\over m^2+(p-K_\|)^2}{2K^-\over K^+ K^2}
-2i\pi\sum_{0<K^+<p^+}{1\over K^{+2}}\Bigg]
\eea
Because of our choice of closed contours, only the pole
at $K^2-i\epsilon=0$ contributes to the integral and then only
when $0<K^+<p^+$:
\bea
-i\Sigma(p)
&=&{ig^2N_c\gamma^+\over8\pi^3}\sum_{0<K^+<p^+}{1\over K^{+2}}\int{d{\bfs K}}
e^{-\delta({\bfs K}+{\bfs k}_0)^2}
{2K^+p^--m^2K^+/(p^+-K^+)\over{\bfs K}^2-2K^+p^- + m^2K^+/(p^+-K^+)}
\nonumber\\
&\sim&{ig^2N_c\gamma^+\over8\pi^2}\sum_{0<K^+<p^+}
\left({m^2+p^2\over p^+K^+}+{m^2\over p^+(p^+-K^+)}\right)
\ln\delta e^\gamma
{K^+(m^2+p^2+2K^+p^-)\over p^+-K^+}\nonumber\\
&\sim&{ig^2N_c\gamma^+\over8\pi^2}\sum_{0<K^+<p^+}
{m^2\over p^+(p^+-K^+)}\ln
{K^{+2}m^2\delta e^\gamma\over p^+(p^+-K^+)}\nonumber\\
&&+(m^2+p^2){ig^2N_c\gamma^+\over8\pi^2}\sum_{0<K^+<p^+} 
{1\over p^+K^+}\ln
{K^{+2}m^2\delta e^{\gamma+1}\over p^+(p^+-K^+)} + O\left([m^2+p^2]^2\right)
\label{smalldelta}
\eea
where the last line applies near mass shell
$m^2+p^2\sim0$.
Stripping away the $\gamma^+$ and multiplying by $2ip^+$
gives the shift in the quantity $m^2+p^2$, and so we find
\bea
\Delta m^2 &=& -{g^2N_c\over4\pi^2}\sum_{0<K^+<p^+}
{m^2\over p^+-K^+}\ln
{K^{+2}m^2\delta e^\gamma\over p^+(p^+-K^+)}\nonumber\\
Z_2&=&1+{g^2N_c\over4\pi^2}\sum_{0<K^+<p^+} 
{1\over K^+}\ln
{K^{+2}m^2\delta e^{\gamma+1}\over p^+(p^+-K^+)}
\eea
The mass shift should be a numerical function of the
UV cutoff $\delta$, but the divergent sum near $K^+=p^+$
introduces an apparent infrared divergence depending on $p^+$.
This is due to the small $\delta$ approximation used in the
second line of (\ref{smalldelta}), which implicitly neglected 
$\delta$ in comparison to the
$p^+$ discretization unit $\epsilon/p^{+2}$. 
If we go back to the on-shell limit of the
first line, we see that the continuum limit of the
$K^+$ sum is actually convergent at fixed $\delta$. 
\bea
-i\Sigma(p)|_{p^2=-m^2}
&=&-{ig^2N_c\gamma^+\over8\pi^3}\sum_{0<K^+<p^+}{m^2\over p^+(p^+-K^+)}
\int{d{\bfs K}}
{e^{-\delta({\bfs K}+{\bfs k}_0)^2}\over{\bfs K}^2+m^2K^{+2}/p^+(p^+-K^+)}
\nonumber\\
&\to&-{ig^2N_cm^2\gamma^+\over8\pi^3p^+}\int_0^1{dx\over 1-x}
\int{d{\bfs K}}
{e^{-\delta({\bfs K}+{\bfs k}_0)^2}\over{\bfs K}^2+m^2x^2/(1-x)}\\
\Delta m^2
&=&{g^2N_cm^2\over4\pi^2}\int_0^\infty{dT\over T+\delta}
\int_0^1{dx\over 1-x}\exp\left\{-{m^2Tx^2\over1-x}-{T\delta{\bfs k}_0^2
\over T+\delta}\right\}
\eea
One can easily check the large and small $T$ behavior of the
function
\bea
F(T)&=&\int_0^1{dx\over 1-x}\exp\left\{-{m^2Tx^2\over1-x}\right\}
\sim\cases{{1\over2}\sqrt{\pi\over m^2T}& {\rm for}~ $T\to\infty$\cr
&\cr
-\ln(m^2Te^\gamma)& {\rm for}~ $T\to0$\cr}
\eea
which confirms that $\Delta m^2$ is finite at fixed $\delta$. Furthermore,
the small $T$ behavior of $F$ controls the small $\delta$ behavior of
$\Delta m^2$:
\bea
\Delta m^2
&\sim&{g^2N_cm^2\over8\pi^2}\left(\ln^2(m^2\delta e^\gamma)
+O(1)\right)
\label{contsmalldelta}
\eea
Comparing this to (\ref{smalldelta}), we that the double logarithmic
UV divergence in (\ref{contsmalldelta}) shows up as a single log
UV times a single log IR divergence in (\ref{smalldelta}) when
$\delta\to0$ is taken before the continuum limit. This nonuniformity
is because in the
latter case the part of the UV divergence due to the zero
thickness of the 1-brane is cut off by the $p^+$ discretization.
However, for $Z_2$ and
the more complicated diagrams considered later, it is valid
to make the small $\delta$ approximation at discrete $p^+$,
because in those cases singularities due to the zero thickness
are integrable. 
The double log  divergence in $\Delta m^2$ would not be 
present if the branion had not
been confined to a brane. Indeed, for a p-brane with $p>1$ the corresponding
singularity would be integrable\footnote{In Feynman gauge 
the loop integral with Dirac fermions on a $p$-brane would be 
in $D$ space-time dimensions
\bea
&&\int {d^DK\over(2\pi)^D}{m-(p-1)\gamma\cdot(p_\|-K_\|)
\over K^2(m^2+(p-K_\|)^2)}
=\int {d^{D-p-1}{\bfs K}d^{p+1}K_\|\over(2\pi)^D}
{m-(p-1)\gamma\cdot(p_\|-K_\|)
\over ({\bfs K}^2+K_\|^2)(m^2+(p-K_\|)^2)}\nonumber\\
&&\to{\Gamma(2-D/2)\over(4\pi)^{D/2}}m^{D-3}{\Gamma(D-3)\Gamma((p+3-D)/2)
\over\Gamma((p-3+D)/2)}\left[1+(p-1){p+3-D\over p-3+D}\right]
\eea 
where in the last line we put the branion on-shell $\gamma\cdot p_\|\to-m$.
For $D\to4$ and $p\to1$, 
we see clearly the double log divergence in the regime
${\bfs K}^2\gg K_\|^2\gg m^2$. 
It occurs because ${\bfs K}^2$ is absent from the
second denominator: it is a branion propagator not
a bulk one. Note that for a p-brane with $p>1$,
the extra divergence would be absent.}.

It is instructive to compare the transverse momentum integral
using our $\delta$ regulator with that using dimensional regularization,
with transverse dimension $d<2$, which gives
\bea
-i\Sigma(p)
&=&{ig^2N_c\gamma^+\over2\pi}\sum_{0<K^+<p^+}{1\over K^{+2}}\int{d{\bfs K}
\over (2\pi)^d}
{2K^+p^--m^2K^+/(p^+-K^+)\over{\bfs K}^2-2K^+p^- + m^2K^+/(p^+-K^+)}
\nonumber\\
&=&-{ig^2N_c\gamma^+\over2\pi}\sum_{0<K^+<p^+}{1\over K^{+2}}{\Gamma(1-d/2)
\over (4\pi)^{d/2}}\left[{m^2K^{+2}\over p^+(p^+-K^+)}+(m^2+p^2){K^+
\over p^+}
\right]^{d/2}
\nonumber\\
&=&-{ig^2N_c\gamma^+\over2\pi}{\Gamma(1-d/2)
\over (4\pi)^{d/2}}
\sum_{0<K^+<p^+}{K^{+ d-2}m^d\over
p^{+d/2}(p^+-K^+)^{d/2} }\left[1+{d\over2}(m^2+p^2){p^+-K^+\over
m^2K^+}
\right]
\eea
where in the last line we expanded about mass shell $p^2+m^2\sim0$,
from which we read off $\Delta m^2$:
\bea
\Delta m^2 &=&{g^2N_c\over\pi}{\Gamma(1-d/2)
\over (4\pi)^{d/2}}
\sum_{0<K^+<p^+}{K^{+ d-2}m^d\over
p^{+d/2-1}(p^+-K^+)^{d/2} }\nonumber\\
&\to&{g^2N_c\over\pi}{m^d\Gamma(1-d/2)
\over (4\pi)^{d/2}}\int_0^1 dx x^{d-2}(1-x)^{-d/2}
={g^2N_c\over\pi}{m^d\Gamma(1-d/2)^2
\over (4\pi)^{d/2}}{\Gamma(d-1)\over\Gamma(d/2)}
\eea
which is finite for $d<2$.
Here the double pole at $d=2$ reflects the double log divergence
in (\ref{contsmalldelta}).
Reading off $Z_2$ we find:
\bea
Z_2&=&1-{g^2N_c\over\pi}{d\over2}{\Gamma(1-d/2)
\over (4\pi)^{d/2}}
\sum_{0<K^+<p^+}{K^{+ d-3}m^{d-2}\over
p^{+d/2-1}(p^+-K^+)^{d/2-1}}\nonumber\\
&\sim&1-{g^2N_c\over4\pi^2}\sum_{0<K^+<p^+}{1\over K^+}
\left[{\Gamma(1-d/2)\over (4\pi)^{(d-2)/2}}-\ln{K^{+2}m^{2}e\over
p^{+}(p^+-K^+)}\right]
\eea
where in the last line we have taken $d\sim 2$. As mentioned above
the singularity for $K^+\to p^+$ is integrable at $d=2$ here, so this
is a valid procedure. We leave $K^+$ discrete because these expressions
are divergent for $d<2$. Comparing to 
the $\delta$ regulator result for $Z_2$ (\ref{smalldelta}), 
we find the correspondence
\bea
{\Gamma(1-d/2)\over (4\pi)^{(d-2)/2}}\leftrightarrow
-\ln e^\gamma\delta\qquad{\rm or}\qquad {2\over2-d}
\leftrightarrow -\ln(4\pi\delta) 
\label{dimdelta}
\eea
Using this correspondence we shall find that in different
diagrams the two regulators are not in precise agreement as 
to the $\delta$ independent terms, and
counterterms must be introduced to achieve equivalent results.
Since dimensional regularization preserves more symmetry
than the $\delta$ regulator, we shall presume that it
is the latter that requires the counterterms. Then simple
comparison of the two regulators in each diagram gives an
efficient procedure for the
identification of the required counterterms.
\subsection{Gluon Self Energy}
From \cite{thornfreedom,chakrabartiqt1} the gluon self energy 
diagram is given 
by\footnote{We correct an error in Eqs. (2.2) 
and (2.12) of \cite{thornfreedom} which
gave -2 instead of $-4/9$ for the constant in $\Pi^{++}$.}
\bea
\Pi^{++}(Q)&\equiv&Q^{+2}\Pi_1\ 
=\ -{g^2N_c\over4\pi^2}Q^{+2}\left({1\over6}\ln\{Q^2\delta e^\gamma\}
-{4\over9}\right)\\
\Pi^{\wedge\vee}(Q)&\equiv&-Q^2\Pi_2\ 
=\ {g^2N_c\over4\pi^2}Q^2\left[{\cal A}(Q^2,Q^+)
-{11\over6}\ln\{Q^2\delta e^\gamma\}+{67\over18}\right]\\
{\cal A}(Q^2,Q^+)&\equiv&\sum_{q^+}\left[{1\over q^+}
+{1\over Q^+-q^+}\right]\ln\left\{x(1-x)Q^2\delta e^\gamma\right\}.
\eea
The corresponding gluon propagator up to one loop is
\bea
D^{--}(Q)&=&{i\over Q^{+2}}\left(
1-\Pi_1\right)\nonumber\\
D^{ij}(Q)&=&{-i\delta_{ij}\over Q^2}\left(
1-\Pi_2\right)
\eea
The contribution of $\Pi$ to one loop branion scattering is then
\bea
{\cal M}_{\rm SE}
&=&{ig^2\gamma_1^+\gamma_2^+}
\left[-{{\bfs Q}^2\Pi_2\over Q^{+2}Q^2}+{\Pi_1\over Q^{+2}}\right]
={ig^2\gamma_1^+\gamma_2^+}
\left[-{2Q^-\Pi_2\over Q^{+}Q^2}+{\Pi_1-\Pi_2\over Q^{+2}}\right]
\nonumber\\
&=&{ig^4N_c\gamma_1^+\gamma_2^+\over4\pi^2}
\Bigg[{2Q^-\over Q^{+}Q^2}\left({\cal A}(Q^2,Q^+)
-{11\over6}\ln\{Q^2\delta e^\gamma\}+{67\over18}\right)\nonumber\\
&&+{1\over Q^{+2}}\left({\cal A}(Q^2,Q^+)
-2\ln\{Q^2\delta e^\gamma\}+{25\over6}\right)\Bigg]
\label{seamp}
\eea
We note that if the integrals are done in dimensional regularization
and the correspondence (\ref{dimdelta}) is assumed then $\Pi_2$
is unchanged and the pure number in $\Pi_1$ replaced as follows
\beq
-{4\over9}\to -{5\over18}
\eeq
that is, $1/6$ is added, so the 25/6 in (\ref{seamp}) is changed
to 4.

\section{Triangle Diagrams}
\label{TriangleSection}
\begin{figure}[ht]
\begin{center}
\psfrag{'k0'}{${\bfs k}_0$}
\psfrag{'k1'}{${\bfs k}_1$}
\psfrag{'l'}{${\bfs l}$}
\psfrag{'q'}{$q$}
\psfrag{'p'}{${p}$}
\psfrag{'pQ'}{$p+Q_\|$}
\psfrag{'qQ'}{$q-Q_\|$}
\psfrag{'K'}{${K}$}
\psfrag{'Q'}{$Q$}
\psfrag{'KQ'}{$Q+K$}
\includegraphics[height=6cm]{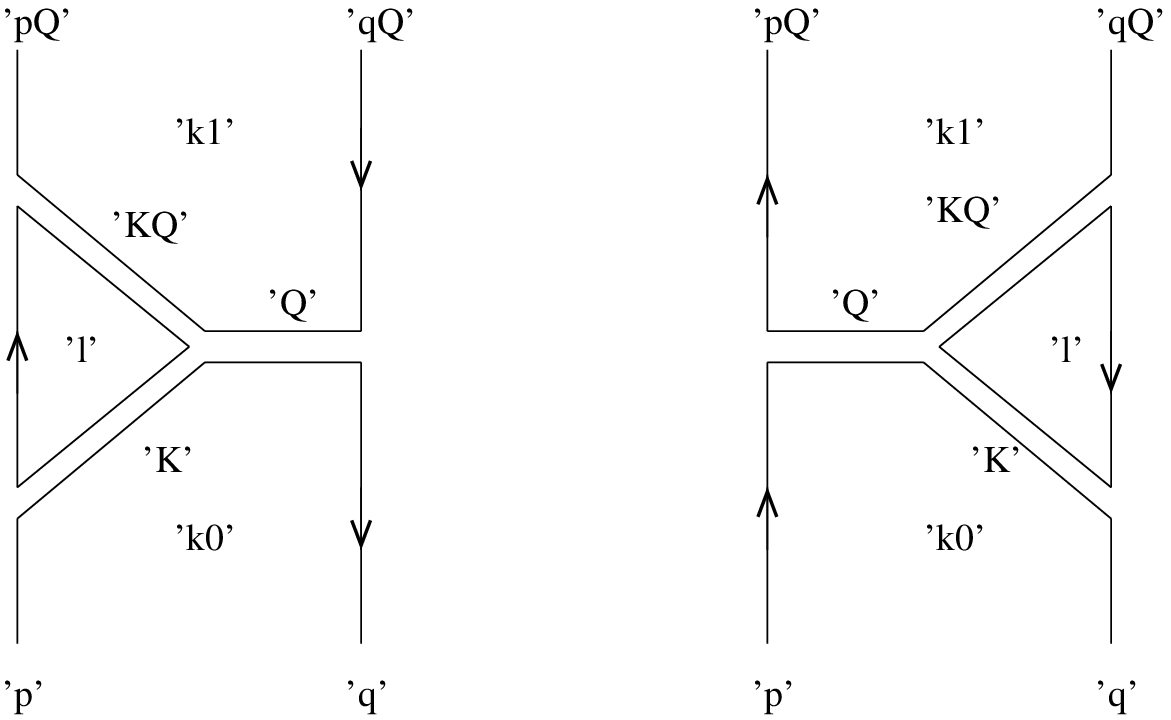}
\end{center}
\caption{Triangle Feynman diagrams contributing to the four-point
amplitude. The arrows show the direction of color flow, and $p,q$ are
incoming momenta.}
\label{TriangleGraphs}
\end{figure}

We now calculate the triangle graphs contributing to the four-point
amplitude. The Feynman diagrams for these contributions are portrayed
in Fig.~\ref{TriangleGraphs}.
%\iffalse
%\fi
Using the light-cone Feynman rules %of Fig.~\ref{FeynmanRules} 
we immediately write the Feynman integral corresponding to the diagram on
the left of Fig.~\ref{TriangleGraphs}.
\begin{eqnarray}
\Gamma_{\triangle_L} &=& N_c \int\!{d^4K\over(2\pi)^4}
(ig\gamma_1^+){-i\over \gamma_1^\alpha(p-K_\|)_\alpha +m}(ig\gamma_1^+)
D^{-\mu_1}(K)D^{-\mu_2}(K+Q)D^{-\mu_3}(Q) \nonumber \\
&& \hskip 3.5cm \times
V_{\mu_1\mu_2\mu_3}(K,-K-Q,Q) (ig\gamma_2^+) \nonumber \\
&\to& -{4g^4N_c\gamma_1^+\gamma_2^+\over Q^+Q^2}{1\over 2\pi}\sum_{K^+}\int\!
{d{\bf K}\over (2\pi)^2}{dK^-\over 2\pi}
{(p^+-K^+) \times F \over 
((p\!-\!K_\|)^2+m^2) K^+K^2 (K^+\!+\!Q^+)(K\!+\!Q)^2},
\label{Triangle1}
\end{eqnarray}
where
\begin{equation}
F = K^-[{\bf Q}^2(2Q^+\!+\!K^+)+{\bf K}\cdot{\bf Q}(2K^+\!+\!Q^+)]
-Q^-[{\bf K}^2(2K^+\!+\!Q^+)+{\bf K}\cdot{\bf Q}(2Q^+\!+\!K^+)].
\label{Triangle2}
\end{equation}
The subscripts on the $\gamma$'s distinguish between the different
branes. All branion-branion-gluon vertices only include the $+$
component of $\gamma^\mu$, since the gluon propagator, $D^{\mu\nu}$,
vanishes when $\mu=+$. We have replaced the $K^+$ integral by
a sum over discretized $K^+=\ell\epsilon$, where $\sum_{K^+}$
means $\epsilon\sum_l$. for each propagator in a
Feynman diagram, the $p^+=0$ term in the sum is excluded.
This discretization and zero mode exclusion serves two purposes: first,
it systematically regulates the artificial $p^+=0$ divergences
that crop up in light-cone gauge, and secondly, it provides
a cutoff to regulate the physical infrared divergences due to
massless gauge particles. In principle, we only take the
continuum $K^+$ limit for properly defined physical quantities.

The diagram on the right of Fig.~\ref{TriangleGraphs} similarly leads
to the integral
\begin{eqnarray}
\Gamma_{\triangle_R} &=& N_c \int\!{d^4K\over(2\pi)^4}
(ig\gamma_2^+){-i\over \gamma_1^\alpha(-q-K_\|)_\alpha +m}(ig\gamma_2^+)
D^{-\mu_1}(K)D^{-\mu_2}(K+Q)D^{-\mu_3}(Q) \nonumber \\
&& \hskip 3.5cm \times
V_{\mu_1\mu_3\mu_2}(-K,-Q,K+Q) (ig\gamma_1^+) \nonumber \\
&\to& -{4g^4N_c\gamma_1^+\gamma_2^+\over Q^+Q^2}{1 \over 2\pi}\sum_{K^+} 
\int\!{d{\bf K}\over (2\pi)^2}{dK^-\over 2\pi}
{(-q^+-K^+) \times F \over 
((q\!+\!K_\|)^2+m^2) K^+K^2 (K^+\!+\!Q^+)(K\!+\!Q)^2},
\label{Triangle1R}
\end{eqnarray}
with the same expression for $F$ given by (\ref{Triangle2}). 

We shall employ dual momentum variables ${\bfs l}, {\bfs k}_0,
{\bfs k}_1$ related to the momenta via
${\bfs Q}={\bfs k}_0-{\bfs k}_1$, ${\bfs K}={\bfs l}-{\bfs k}_0$.
We shall specify our ultraviolet cutoff in these variables, by
supplying a factor $e^{-\delta{\bfs l}^2}$.
When the branions are on-shell, we can use the identity
\bea
2(p^+-K^+)K^-&=&m^2+(p-K_\|)^2-{m^2K^+\over p^+}\qquad {\rm for}~\triangle_L
\label{iden1}\\
2(-q^+-K^+)K^-&=&m^2+(q+K_\|)^2+{m^2K^+\over q^+}\qquad {\rm for}~\triangle_R
\label{iden1R}\eea
to rewrite the $K^-$ term in $F$. Substituting (\ref{iden1}) or (\ref{iden1R})
into (\ref{Triangle1}) or (\ref{Triangle1R}), 
we see that the first two terms on the
right of either identity cancel a propagator, contributing a bubble-like
integral to the vertex function:
\bea
\hskip-10pt
\Gamma^{\rm Bubble}_{\triangle_L}&=&\Gamma^{\rm Bubble}_{\triangle_R}
\nonumber\\
&=&-{2g^4N_c\gamma_1^+\gamma_2^+\over Q^+Q^2}{1\over2\pi}\sum_{K^+}\int\!
{d{\bf K}\over (2\pi)^2}{dK^-\over 2\pi}
{{\bf Q}^2(2Q^+\!+\!K^+)+{\bf K}\cdot{\bf Q}(2K^+\!+\!Q^+) \over 
K^+K^2 (K^+\!+\!Q^+)(K\!+\!Q)^2}\nonumber\\
&=&-{g^4N_c\gamma_1^+\gamma_2^+\over Q^+Q^2}{i{\rm sgn}Q^+\over
(2\pi)^3 }\sum_{K^+}\int\!{d{\bf K}}{e^{-\delta({\bf K}+{\bf k}_0)^2}
\over Q^+K^+(Q^++K^+)}
{{\bf Q}^2(2Q^+\!+\!K^+)+{\bf K}\cdot{\bf Q}(2K^+\!+\!Q^+) \over 
({\bf K}+x{\bf Q})^2+x(1-x)Q^2}
\label{bubbleminusint}
\eea
where we have put $x=-K^+/Q^+$, and have done 
$\int dK^-$ by a contour chosen to pick up the
$K^2$ pole, closing in the 
upper half plane for $Q^+>0$ and in the lower half plane for $Q^+<0$.
A pole is enclosed by the contour only when $0<-K^+/Q^+<1$.
In the rest of this paper we shall for simplicity and
definiteness assume $Q^+>0$.
We have also inserted the worldsheet friendly cutoff $e^{-\delta{\bfs l}^2}$ 
on the dual loop momentum variable ${\bfs l}={\bfs K}+{\bfs k}_0$. 
To do the transverse momentum integral,
we exponentiate the denominator with a Schwinger representation
$1/D=\int_0^\infty dT e^{-DT}$ and complete the square
in the exponent:
\bea
(\delta+T){\bf K}^2+2{\bf K}\cdot(\delta{\bf k}_0+xT{\bf Q})+\delta{\bf k}_0^2
+x^2T{\bf Q}^2=(\delta+T)\left({\bf K}+{\delta{\bf k}_0
+xT{\bf Q}\over\delta+T}\right)^2+{\delta T({\bf k}_0-x{\bf Q})^2
\over \delta+T}
\eea
Then doing the Gaussian integral, this term becomes
\bea
&&\hskip-1in-{g^4N_c\gamma_1^+\gamma_2^+\over Q^+Q^2}{i\over
8\pi^2 }\sum_{K^+}\int\!{dT\over T+\delta}{\exp\{-Tx(1-x)Q^2-\delta 
T({\bf k}_0-x{\bf Q})^2/(T+\delta)\}
\over Q^+K^+(Q^++K^+)}\nonumber\\
&&\hskip+1in\times\left({{\bf Q}^2(2Q^+\!+\!K^+)-\left[{\delta{\bf k}_0
+xT{\bf Q}\over\delta+T}\right]\cdot{\bf Q}(2K^+\!+\!Q^+)}\right)\nonumber\\
&=&{g^4N_c\gamma_1^+\gamma_2^+\over Q^{+3}Q^2}{i\over
8\pi^2 }\sum_{K^+}\int\!{dT\over T+\delta}{\exp\{-Tx(1-x)Q^2-\delta 
T({\bf k}_0-x{\bf Q})^2/(T+\delta)\}
\over x(1-x)}\nonumber\\
&&\hskip1in\times\left({{\bf Q}^2(2-x)-\left[{\delta{\bf k}_0
+xT{\bf Q}\over\delta+T}\right]\cdot{\bf Q}(1-2x)}\right)
\eea
Some integrals:
\bea
\int_0^\infty{dT\over T+\delta}e^{-TA}&\sim& \Gamma^\prime(1)
-\ln(A\delta)\ =\ -\ln(A\delta e^\gamma)\nonumber\\
\quad \int_0^\infty{TdT\over (T+\delta)^2}
e^{-TA}&\sim& -\ln(A\delta e^{\gamma+1}), \qquad\qquad
\int_0^\infty{dT\delta\over (T+\delta)^2}e^{-TA}\sim 1
\eea
in the limit $\delta\to0$, where we have introduced Euler's constant $\gamma
=-\Gamma^\prime(1)$. Then
%\iffalse
\bea
\Gamma^{\rm Bubble}_{\triangle_L}&=&
{g^4N_c\gamma_1^+\gamma_2^+\over Q^{+3}Q^2}{i\over
8\pi^2 }\sum_{K^+}{1\over x(1-x)}
\bigg({\bf Q}^2\left[\left\{(2-x)-x(1-2x)\right\}
\left(-\ln(x(1-x)Q^2\delta e^\gamma)\right)\right]\nonumber\\
&&+x(1-2x){\bf Q}^2-{\bf k}_0\cdot{\bf Q}(1-2x)\bigg)\nonumber\\
&=&{g^4N_c\gamma_1^+\gamma_2^+\over Q^{+3}Q^2}
{i\over8\pi^2 }\sum_{K^+}
\bigg({\bf Q}^2\left[{2(1-x(1-x))\over x(1-x)}
\left(-\ln(x(1-x)Q^2\delta e^\gamma)\right)
-{(1-2x)^2\over 2x(1-x)}\right]
\nonumber\\
&&-({\bf k}^2_0-{\bf k}_1^2){1-2x\over 2x(1-x)}\bigg)\nonumber\\
&=&{g^4N_c\gamma_1^+\gamma_2^+\over Q^{+3}Q^2}
{i\over8\pi^2 }\sum_{K^+}
{\bf Q}^2\left[{2(1-x(1-x))\over x(1-x)}
\left(-\ln(x(1-x)Q^2\delta e^\gamma)\right)
-{(1-2x)^2\over 2x(1-x)}\right]\nonumber\\
&\to&{g^4N_c\gamma_1^+\gamma_2^+\over Q^{+3}Q^2}
{i\over4\pi^2 }{\bf Q}^2\left[\sum_{K^+}
{1\over x(1-x)}
\left(-\ln(x(1-x)Q^2\delta e^\gamma)-{1\over4}\right)
+Q^+\left(\ln(Q^2\delta e^\gamma)-1\right)\right]\nonumber\\
&\to&-{ig^4N_c\gamma_1^+\gamma_2^+{\bfs Q}^2\over4\pi^2  Q^{+2}Q^2}
\left[{\cal A}(Q^2,Q^+)
+{1\over4Q^+}\sum_{K^+}{1\over x(1-x)}
-\ln(Q^2\delta e^\gamma)+1\right]
\label{trianglebubble}
\eea
%\fi
where we have used ${\bf Q}={\bf k}_0-{\bf k}_1$. Note that the last 
term of the second equality,
 which depends on the ${\bf k}$'s individually vanishes after summation
on $K^+$ because the summand is odd under $x\to(1-x)$. We recall that the
bubble contribution from the right triangle diagram is identical to this
\bea
\Gamma^{\rm Bubble}_{\triangle_R}&=&\Gamma^{\rm Bubble}_{\triangle_L}
\eea

The dimensional regularization evaluation of the bubble integral
(\ref{bubbleminusint}) is very simple (recall $Q^+>0$)
\bea
\Gamma^{\rm Bubble}_{\triangle_L}
&=&-{ig^4N_c\gamma_1^+\gamma_2^+\over 2\pi Q^+Q^2}\sum_{K^+}
\int\!{d{\bf K}\over(2\pi)^d}{1
\over Q^+K^+(Q^++K^+)}
{{\bf Q}^2(2Q^+\!+\!K^+)+{\bf K}\cdot{\bf Q}(2K^+\!+\!Q^+) \over 
({\bf K}+x{\bf Q})^2+x(1-x)Q^2}\nonumber\\
&=&{ig^4N_c\gamma_1^+\gamma_2^+{\bf Q}^2\over 4\pi^2 Q^{+2}Q^2}\sum_{K^+}
{\Gamma(1-d/2)\over(4\pi)^{(d-2)/2}}{(x(1-x)Q^2)^{(d-2)/2}
\over Q^+x(1-x)}
[1-x(1-x)]\nonumber\\
&\sim&{ig^4N_c\gamma_1^+\gamma_2^+{\bf Q}^2\over 4\pi^2 Q^{+2}Q^2}\sum_{K^+}
{1-x(1-x)\over Q^+x(1-x)}
\left[{\Gamma(1-d/2)\over(4\pi)^{(d-2)/2}}-\ln(x(1-x)Q^2)
\right]\nonumber\\
&\sim&{ig^4N_c\gamma_1^+\gamma_2^+{\bf Q}^2\over 4\pi^2 Q^{+2}Q^2}\sum_{K^+}
{1-x(1-x)\over Q^+x(1-x)}
\left[-\ln(x(1-x)Q^2\delta e^\gamma)\right]\nonumber\\
&=&-{ig^4N_c\gamma_1^+\gamma_2^+{\bf Q}^2\over 4\pi^2 Q^{+2}Q^2}\sum_{K^+}
\left[{\cal A}(Q^2, Q^+) - \ln(Q^2\delta e^\gamma)+2\right]
\label{bubbleintdimreg}
\eea
We see that the dim-reg evaluation doesn't show the 
second term in square brackets of the $\delta$ evaluation. We shall
see later that this term would spoil the cancellation of
infrared divergences and should in fact be absent. So we
identify it as a term to be cancelled by a counterterm.

The rest of each triangle diagram involves all three propagators,
but in $F$ the factor $K^-$ is replaced by $-m^2K^+/2p^+(p^+-K^+)$
for the left triangle and by $-m^2K^+/2q^+(q^++K^+)$ for the right triangle.
Thus the two numerators are replaced as follows
\bea
2(p^+-K^+)F&\to&
-{m^2K^+\over p^+}[{\bf Q}^2(2Q^+\!+\!K^+)+{\bf K}\cdot{\bf Q}(2K^+\!+\!Q^+)]
\nonumber\\
&&+(p^+-K^+)
{Q_\|^2\over Q^+}[{\bf K}^2(2K^+\!+\!Q^+)+{\bf K}\cdot{\bf Q}(2Q^+\!+\!K^+)]
\\
-2(q^++K^+)F&\to&
{m^2K^+\over q^+}[{\bf Q}^2(2Q^+\!+\!K^+)+{\bf K}\cdot{\bf Q}(2K^+\!+\!Q^+)]
\nonumber\\
&&-(q^++K^+)
{Q_\|^2\over Q^+}[{\bf K}^2(2K^+\!+\!Q^+)+{\bf K}\cdot{\bf Q}(2Q^+\!+\!K^+)]
\eea
The integration over $K^-,{\bfs K}$ is evaluated in Appendices A, B. The
$K^-$ integration restricts the range of $K^+$ to two distinct regions
for each triangle diagram. Then the transverse can have three distinct
numerators $1$, ${\bfs K}^2$, and ${\bfs K}\cdot{\bfs Q}$. In the notation of 
the appendices we then have for the left triangle
\bea
\Gamma^{\rm Rest}_{\triangle_L}&=&-{g^4N_c\gamma_1^+\gamma_2^+\over
\pi Q^+Q^2}\int_0^{p^+}{dK^+\over K^+(K^++Q^+)}
\bigg(-{m^2K^+\over p^+}\left[{\bf Q}^2(2Q^+\!+\!K^+)I^1_L
+(2K^+\!+\!Q^+)I^1_L[{\bf K}\cdot{\bf Q}]\right]
\nonumber\\
&&\qquad +(p^+-K^+)
{Q_\|^2\over Q^+}\left[(2K^+\!+\!Q^+)I^1_L[{\bf K}^2]
+(2Q^+\!+\!K^+)I^1_L[{\bf K}\cdot{\bf Q}]\right]
\bigg)\nonumber\\
&&-{g^4N_c\gamma_1^+\gamma_2^+\over
\pi Q^+Q^2}\int_{-Q^+}^{0}{dK^+\over K^+(K^++Q^+)}
\bigg(-{m^2K^+\over p^+}\left[{\bf Q}^2(2Q^+\!+\!K^+)I^2_L
+(2K^+\!+\!Q^+)I^2_L[{\bf K}\cdot{\bf Q}]\right]
\nonumber\\
&&\qquad+(p^+-K^+)
{Q_\|^2\over Q^+}\left[(2K^+\!+\!Q^+)I^2_L[{\bf K}^2]
+(2Q^+\!+\!K^+)I^2_L[{\bf K}\cdot{\bf Q}]\right]
\bigg)
\eea
We have written the $K^+$ sums as continuous integrals, because
inspection of the tables of asymptotics in Appendix B shows
that the potential divergences due to the factors $1/K^+(K^++Q^+)$
are absent: The singularity at $K^+=0$ is integrable because 
the coefficient of $1/K^+$ is continuous through $K^+=0$
and the continuum limit of the sum leads to the principal value
prescription The singularity at $K^+=-Q^+$ is 
integrable because the coefficient of $1/(K^++Q^+)$ vanishes
as $K^+\to -Q^+$.
However, these integrals do have some residual $\delta$
dependence. Eqs. (\ref{ksquared1}), (\ref{ksquared2}) of Appendix B 
show that
\bea
I^1_L[{\bfs K}^2]
&\equiv& {\hat I}^1_L[{\bfs K}^2] -{i\over8\pi(p^+-K^+)}\ln(m^2\delta e^\gamma)
\\
I^2_L[{\bfs K}^2]
&\equiv&{\hat I}^2_L[{\bfs K}^2]-{i\over8\pi (p^+-K^+)}{K^++Q^+\over Q^+}
\ln(m^2\delta e^\gamma)
\eea
Where the notation ${\hat X}$ signifies that $\delta e^\gamma$ in
$X$ is replaced by $1/m^2$. Then with this same notation, we
can write
\bea
\Gamma^{\rm Rest}_{\triangle_L}&=&{\hat\Gamma}^{\rm Rest}_{\triangle_L}
+{ig^4N_c\gamma_1^+\gamma_2^+\over
8\pi^2 Q^+Q^2}{Q_\|^2\over Q^+}
\bigg(\int_0^{p^+}{dK^+(2K^+\!+\!Q^+)\over K^+(K^++Q^+)}
+\int_{-Q^+}^{0}{dK^+(2K^+\!+\!Q^+)\over K^+Q^+}
\bigg)\ln(m^2\delta e^\gamma)\nonumber\\
&=&{\hat\Gamma}^{\rm Rest}_{\triangle_L}
+{ig^4N_c\gamma_1^+\gamma_2^+\over
8\pi^2 Q^+Q^2}{Q_\|^2\over Q^+}
\bigg(-\int_0^{p^+}{dK^+(2K^+\!+\!Q^+)\over Q^+(K^++Q^+)}
+-\hskip-10pt\int_{-Q^+}^{p^+}{dK^+(2K^+\!+\!Q^+)\over K^+Q^+}
\bigg)\ln(m^2\delta e^\gamma)\nonumber\\
&=&{\hat\Gamma}^{\rm Rest}_{\triangle_L}
-{ig^4N_c\gamma_1^+\gamma_2^+Q^-\over
4\pi^2 Q^+Q^2}
\left(\ln{p^++Q^+\over Q^+} 
+\ln{p^+\over Q^+}+2 \right)\ln(m^2\delta e^\gamma)
\eea
where the line through the integral sign on the second
line denotes a principal value prescription.
Incidentally, these three lines show explicitly the infrared divergence 
cancellation sketched above, for the $\delta$ dependence.
 
Finally we quote the complete left triangle diagram:
\begin{eqnarray}
\Gamma_{\triangle_L} &=&
{\hat\Gamma}^{\rm Rest}_{\triangle_L}
-{ig^4N_c\gamma_1^+\gamma_2^+\over
4\pi^2 Q^+Q^2}\Bigg\{Q^-
\left(\ln{p^++Q^+\over Q^+} 
+\ln{p^+\over Q^+}+2 \right)\ln(m^2\delta e^\gamma)
\nonumber\\
&&+{{\bf Q}^2\over Q^{+}}
\left[{\cal A} - \ln(Q^2\delta e^\gamma) +2\right]+{{\bf Q}^2\over Q^{+}}
\left[ 
{1\over 4Q^+}\sum_{K^+}{1\over x(1-x)}-1\right]\Bigg\}.
\label{TriangleResult}
\end{eqnarray}
The last term in braces is absent in dim-reg.
Note that the top line is finite in the
infrared (continuous $K^+$).
A similar result corresponding to the Feynman diagram on the
right side of Fig.~\ref{TriangleGraphs} may be obtained either directly 
or from Eq.~(\ref{TriangleResult}) by the substitution\footnote{Strictly 
speaking, we should also substitute ${\bfs k}_{0,1}
\to-{\bfs k}_{1,0}$. But this is not necessary since
Eq.~(\ref{TriangleResult}) only depends on 
the dual momenta through ${\bfs k}_1-{\bfs k}_0$
which is invariant under this last substitution.},
$p\rightarrow q-Q_\|$. Note that the only {\it on-shell} value
of $Q^+>0$ is $q^+-p^+$, so $q^+-Q^+=p^+$. In this case, the right
triangle contribution is precisely the same as the left triangle
contribution:
\bea
\Gamma_{\triangle_R} &=&\Gamma_{\triangle_L} \qquad {\rm On-Shell}
\eea 
We shall also have use for a slight rearrangement of (\ref{TriangleResult})
where we use ${\bfs Q}^2=Q^2+2Q^+Q^-$:
\begin{eqnarray}
\Gamma_{\triangle_L} &=&
{\hat\Gamma}^{\rm Rest}_{\triangle_L}
-{ig^4N_c\gamma_1^+\gamma_2^+\over
4\pi^2 Q^+Q^2}\Bigg\{Q^-\bigg[
\left(\ln{p^++Q^+\over Q^+} 
+\ln{p^+\over Q^+}\right)\ln(m^2\delta e^\gamma)
+2{\cal A} - 2\ln{Q^2\over m^2} +4\bigg]
\nonumber\\&& 
+{Q^2\over Q^{+}}
\bigg[{\cal A} - \ln(Q^2\delta e^\gamma) +2\bigg] 
+{{\bfs Q}^2\over Q^+}\bigg[{1\over 4Q^+}
\sum_{K^+}{1\over x(1-x)}-1\bigg]\Bigg\}.
\label{TriangleResult2}
\end{eqnarray}
At this point we give the triangle combined with the 
wave function renormalization factors $\sum(Z_i-1)/2$ 
associated with the three external legs:
\bea
&&\hskip-.5in\Gamma_{\triangle_L}+{1\over2}M_{\rm SE}+{1\over 2}(
Z_2(p)+Z_2(p+Q)-2){2ig^2\gamma_1^+\gamma_2^+Q^-\over Q^+Q^2} \ =\
\nonumber\\&& 
{\hat\Gamma}^{\rm Rest}_{\triangle_L}
-{ig^4N_c\gamma_1^+\gamma_2^+\over4\pi^2Q^2Q^{+2}}
\Bigg\{-Q^+Q^-\bigg( -\left(\ln{p^++Q^+\over Q^+} 
+\ln{p^+\over Q^+}\right)\ln(m^2\delta e^\gamma)
+ 2\ln{Q^2\over m^2} - 4\nonumber\\
&&\nonumber\\
&&+\sum_{0<K^+<p^+} 
{1\over K^+}\ln
{K^{+2}m^2\delta e^{\gamma+1}\over p^+(p^+-K^+)}
+\sum_{0<K^+<p^++Q^+} 
{1\over K^+}\ln
{K^{+2}m^2\delta e^{\gamma+1}\over (p^++Q^+)(p^++Q^+-K^+)}
\nonumber\\
&&-{\cal A}(Q^2,Q^+)-{11\over6}\ln\{Q^2\delta e^\gamma\}+{67\over18}\bigg)
\nonumber\\
&&+{{Q}^2}
\bigg[{1\over2}{\cal A}(Q^2,Q^+)
-{25\over12}+2\bigg]
+{\bfs Q}^2\bigg[{1\over4Q^+}\sum_{-Q^+<K^+<0}{1\over x(1-x)}-1\bigg]
\Bigg\}\nonumber\\
&=& {\hat\Gamma}^{\rm Rest}_{\triangle_L}
-{ig^4N_c\gamma_1^+\gamma_2^+\over4\pi^2Q^2Q^{+2}}
\Bigg\{-Q^+Q^-\bigg(2\ln{Q^2\over m^2} - 4\nonumber\\
&&+\sum_{0<K^+<p^+} 
{1\over K^+}\ln
{K^{+2}e\over p^+(p^+-K^+)}
+\sum_{0<K^+<p^++Q^+} 
{1\over K^+}\ln
{K^{+2}e\over (p^++Q^+)(p^++Q^+-K^+)}
\nonumber\\
&&-2\sum_{0<K^+<Q^+}{1\over K^+}\ln{K^+(Q^+-K^+)Q^2\over
m^2Q^{+2}}-{11\over6}\ln\{Q^2\delta e^\gamma\}+{67\over18}\bigg)
\nonumber\\
&&+{{Q}^2}
\bigg[{1\over2}{\cal A}(Q^2,Q^+)
-{1\over12}\bigg]
+{\bfs Q}^2\bigg[{1\over4Q^+}\sum_{-Q^+<K^+<0}{1\over x(1-x)}-1\bigg]
\Bigg\}\nonumber\\
&\to& {\hat\Gamma}^{\rm Rest}_{\triangle_L}
-{ig^4N_c\gamma_1^+\gamma_2^+\over4\pi^2Q^2Q^{+2}}
\Bigg\{-Q^+Q^-\bigg(-{11\over6}\ln\{Q^2\delta e^\gamma\}+{67\over18}
+2\ln{Q^2\over m^2} - 4
%\nonumber\\&&
+\sum_{0<K^+<p^+} 
{1\over K^+}\ln
{K^{+2}e\over p^{+2}}\nonumber\\&&
+\sum_{0<K^+<p^++Q^+} 
{1\over K^+}\ln
{K^{+2}e\over (p^++Q^+)^2}
%\nonumber\\&&
-2\sum_{0<K^+<Q^+}{1\over K^+}\ln{K^+Q^2\over
m^2Q^{+}}+{2\pi^2\over 3}
\bigg)
\nonumber\\
&&+{{Q}^2}
\bigg[{1\over2}{\cal A}(Q^2,Q^+)
-{1\over12}\bigg]
+{\bfs Q}^2\bigg[{1\over4Q^+}\sum_{-Q^+<K^+<0}{1\over x(1-x)}-1\bigg]
\Bigg\}\nonumber\\
&\to&{\hat\Gamma}^{\rm Rest}_{\triangle_L}
-{ig^4N_c\gamma_1^+\gamma_2^+\over4\pi^2Q^2Q^{+2}}
\Bigg\{-Q^+Q^-\bigg(-{11\over6}\ln\{Q^2\delta e^\gamma\}+{67\over18}
+2\ln{Q^2\over m^2} - 4 +{2\pi^2\over 3}\nonumber\\
&&+\sum_{0<K^+<Q^+} 
{2\over K^+}\ln
{m^2K^{+}Q^+e\over Q^2p^{+}(p^++Q^+)}
+\ln{p^+\over Q^+}-\ln^2{p^+\over Q^+}+\ln{p^++Q^+\over Q^+}
-\ln^2{p^++Q^+\over Q^+}\bigg)
\nonumber\\
&&+{Q^2\over2}{\cal A}(Q^2,Q^+)-{Q^2\over12}
+{\bfs Q}^2\bigg[{1\over4Q^+}\sum_{-Q^+<K^+<0}{1\over x(1-x)}-1\bigg]
\Bigg\}
\label{trianglepluswf}
\eea
Arrows indicate that some finite $K^+$ sums have been replaced by
integrals and evaluated.
We see that the ultraviolet divergence is that of asymptotic freedom in
the first group of terms multiplying $Q^-$. In the last line there
is still $\delta$ dependence in ${\cal A}$ that will be cancelled by 
a term from the box diagram (see Eq. (\ref{boxbubble1})). 
Also there are uncanceled infrared divergences in
the last two lines. In dimensional regularization, the last term
multiplying ${\bfs Q}^2$ is absent, and also the $-Q^2/12$ in the
last line is absent:
\bea
&&\hskip-.5in\Gamma_{\triangle_L}+{1\over2}M_{\rm SE}+{1\over 2}(
Z_2(p)+Z_2(p+Q)-2){2ig^2\gamma_1^+\gamma_2^+Q^-\over Q^+Q^2}\nonumber\\
&\to& {\hat\Gamma}^{\rm Rest}_{\triangle_L}
-{ig^4N_c\gamma_1^+\gamma_2^+\over4\pi^2Q^2Q^{+2}}
\Bigg\{-Q^+Q^-\bigg(-{11\over6}\ln\{Q^2\delta e^\gamma\}+{67\over18}
+2\ln{Q^2\over m^2} - 4 +{2\pi^2\over 3}\nonumber\\
&&+\sum_{0<K^+<Q^+} 
{2\over K^+}\ln
{m^2K^{+}Q^+e\over Q^2p^{+}(p^++Q^+)}
+\ln{p^+\over Q^+}-\ln^2{p^+\over Q^+}+\ln{p^++Q^+\over Q^+}
-\ln^2{p^++Q^+\over Q^+}\bigg)
\nonumber\\
&&+
{{Q}^2\over2}{\cal A}(Q^2,Q^+)\Bigg\}\hskip2in {\rm Dim-Reg}
\label{trianglepluswfdim}
\eea
We shall assume that dimensional regularization is correct, in which
case the worldsheet friendly $\delta$ regularization counterterms must
be included which produce the contribution
\bea
\Gamma^{\rm C.T.}&=&{ig^4N_c\gamma_1^+\gamma_2^+\over4\pi^2Q^2Q^{+2}}
\Bigg\{-{Q^2\over12}
+{\bfs Q}^2\bigg[{1\over4Q^+}\sum_{-Q^+<K^+<0}{1\over x(1-x)}-1\bigg]
\Bigg\}\\
&=&{ig^4N_c\gamma_1^+\gamma_2^+\over4\pi^2Q^2Q^{+2}}
\Bigg\{2Q^+Q^-\bigg[{1\over4}\sum_{-Q^+<K^+<0}\left[{1\over K^++Q^+}
-{1\over K^+}\right]-1\bigg]\nonumber\\
&&+Q^2\bigg[{1\over4}\sum_{-Q^+<K^+<0}\left[{1\over K^++Q^+}
-{1\over K^+}\right]-{13\over12}\bigg]
\Bigg\}
\label{counterterm}
\eea
This will be a challenge for the worldsheet formalism to reproduce
locally in view of the $1/K^+$ terms. We shall return to this in
the concluding section.

\section{Box Diagram}
\begin{figure}[ht]
\begin{center}
\psfrag{'k0'}{${\bfs k}_0$}
\psfrag{'k1'}{${\bfs k}_1$}
\psfrag{'l'}{${\bfs l}$}
\psfrag{'q'}{$q$}
\psfrag{'p'}{${ p}$}
\psfrag{'pQ'}{$p+Q_\|$}
\psfrag{'qQ'}{$q-Q_\|$}
\psfrag{'K'}{${K}$}
\psfrag{'Q'}{$Q$}
\psfrag{'KQ'}{$Q+K$}
\includegraphics[height=6cm]{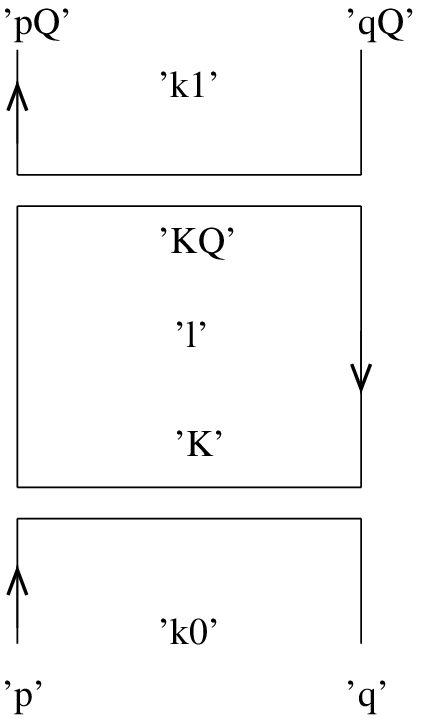}
\end{center}
\caption{Box Feynman diagram contributing to the four-point
amplitude. The arrows show the direction of color flow, and $p,q$ are
incoming momenta.}
\label{Boxgraph}
\end{figure}

Finally we turn to the box diagram drawn in Fig.~\ref{Boxgraph}.
The factors in the numerator of the box integrand can be
written
\bea
-4(p^+-K^+)(q^++K^+)K^-(K^-+Q^-)&=&[m^2+(p-K_\|)^2][m^2+(q+K_\|)^2]
\nonumber\\
&&\hskip-2in-[m^2+(p-K_\|)^2]{m^2(p^+-K^+)\over p^++q^+}-[m^2+(q+K_\|)^2]
{m^2(q^++K^+)\over p^++q^+}
\eea
which shows that the box integration can be reduced to bubble-like and
triangle-like integrations.
\begin{eqnarray}
&&\Gamma_{\Box}=
-{4g^4N_c\gamma_1^+\gamma_2^+}
\int{d^4K\over(2\pi)^4}e^{-\delta({\bfs K}+{\bfs k}_0)^2}
\bigg[{m^2(q^++K^+)\over(p^++q^+)K^+(K^++Q^+)
K^2(K+Q)^2(m^2+(p-K_\|)^2)}\nonumber\\
&&\quad +{m^2(p^+-K^+)\over(p^++q^+)K^+(K^++Q^+)
K^2(K+Q)^2(m^2+(q+K_\|)^2)}-{1\over K^+(K^++Q^+)
K^2(K+Q)^2}\bigg]
\label{gammabox}
\end{eqnarray}
The last term in square brackets of (\ref{gammabox}) 
is the integrand of a bubble 
diagram, whose evaluation is similar to that of 
the corresponding term in
the triangle diagram. The result is, for $Q^+>0$ and $\delta\sim0$,
\bea
\Gamma_{\Box}^{\rm~bubble}=
{ig^4N_c\gamma_1^+\gamma_2^+\over4\pi^2Q^{+3}}\sum_{K^+}{1\over x(1-x)}
\ln\{x(1-x)Q^2\delta e^\gamma\}=
{ig^4N_c\gamma_1^+\gamma_2^+\over4\pi^2Q^{+2}}{\cal A}(Q^2,Q^+)
\label{boxbubble1}
\eea
with $0<x=-K^+/Q^+<1$. We see that half of this
contribution precisely cancels
the ${\cal A}$ term in the last line of (\ref{trianglepluswf})
or (\ref{trianglepluswfdim}).
The other half cancels the corresponding term for a triangle vertex
on the right of the four branion graph. 

The rest of the box diagram is given by two triangle integrands
whose $K^-$ and ${\bfs K}$ integration is given in Appendix A. Since
these contributions are finite in the ultraviolet, we
may set $\delta=0$ in them:
\bea
\Gamma_{\Box}^{\rm~rest}&=&\Gamma_{\Box_L}^{\rm~rest}+
\Gamma_{\Box_R}^{\rm~rest}\\
\Gamma_{\Box_L}&\equiv&
-{4g^4N_c\gamma_1^+\gamma_2^+m^2\over 2\pi(p^++q^+)}
\bigg\{\sum_{0<K^+<p^+}{(q^++K^+)I^1_L\over K^+(K^++Q^+)}+
\sum_{-Q^+<K^+<0}{(q^++K^+)I^2_L\over K^+(K^++Q^+)}\bigg\}\\
\Gamma_{\Box_R}&\equiv&-{4g^4N_c\gamma_1^+\gamma_2^+m^2\over 2\pi(p^++q^+)}
\bigg\{ \sum_{-Q^+<K^+<0}{(p^+-K^+)I^2_R\over K^+(K^++Q^+)} 
+\sum_{-q^+<K^+<-Q^+}{(p^+-K^+)I^1_R\over K^+(K^++Q^+)}\bigg\}
\label{restbox}
\eea
By consulting the asymptotics tables of Appendix B, we see that
there are infrared divergences that prevent immediately converting
these sums to integrals over $K^+$. However, we can neatly extract 
the divergent structure by removing the asymptotic
forms from each of the $I$'s. This can be done in many ways. Introducing a
parameter $\xi$, we define
\bea
I^1_L&\equiv& {\check I}^1_L(\xi) 
+ \xi{i(Q^++K^+)\over8\pi p^+ Q^+Q^2}\ln{Q^4p^{+2}\over m^2Q_\|^2K^{+2}}\\
I^2_L&\equiv& {\check I}^2_L(\xi) 
+ \xi{i(Q^++K^+)\over8\pi p^+ Q^+Q^2}\ln{Q^4p^{+2}\over m^2Q_\|^2K^{+2}}
\nonumber\\
&&- {i(Q^++K^+)\over8\pi Q^+ p^+ Q^2}\ln{Q^2p^{+2}\over m^2Q^+(-K^{+})}
-{iK^+\over8\pi Q^+(p^++Q^+) Q^2}
\ln{Q^2(p^++Q^+)^2\over m^2Q^+(K^++Q^+)}\\
I^1_R&\equiv& {\check I}^1_R(\xi)
- \xi{iK^+\over8\pi (q^+-Q^+)Q^+Q^2}
\ln{Q^4(q^{+}-Q^+)^2\over m^2Q_\|^2(K^{+}+Q^+)^2}
\\
I^2_R&\equiv& {\check I}^2_R(\xi) 
- \xi{iK^+\over8\pi (q^+-Q^+)Q^+Q^2}
\ln{Q^4(q^{+}-Q^+)^2\over m^2Q_\|^2(K^{+}+Q^+)^2}
\nonumber\\
&&+ {iK^+\over8\pi Q^+(q^+-Q^+) Q^2}
\ln{Q^2(q^{+}-Q^+)^2\over m^2Q^+(K^{+}+Q^+)}
+ {i(Q^++K^+)\over8\pi Q^+(q^+Q^2)}\ln{Q^2q^{+2}\over m^2Q^+(-K^+)}
\eea
where we have used ${\check X}$ to denote the part of $X$ that
leads to IR finite integrals over $K^+$. Notice that in the cases
$I^2_{L,R}$ when
both endpoints give divergences, we have multiplied the corresponding
asymptotic forms by factors that are unity at the corresponding
endpoint and vanish at the opposite endpoint. We included the
corresponding factors in $I^1_{L,R}$ for reasons of continuity.
Symmetric summation about the interior singular point shows that
the ${\check I}(\xi)$'s are free of IR divergences for all $\xi$.
The case $\xi=1$ separates the divergent pieces of each ${\check I}$.
However we shall hereafter choose $\xi=0$, to keep subsequent
expressions as simple as possible. Then we define ${\check I}\equiv
{\check I}(0)$:
\bea
\Gamma_{\Box_L}&\equiv&{\check\Gamma}_{\Box_L}
-{ig^4N_c\gamma_1^+\gamma_2^+m^2\over 4\pi^2(p^++q^+)Q^+Q^2}
\bigg\{-\sum_{-Q^+<K^+<0}{(q^++K^+)\over K^+ p^+}
\ln{Q^2p^{+2}\over m^2Q^+(-K^{+})}\nonumber\\
&&-\sum_{-Q^+<K^+<0}{(q^++K^+)\over (K^++Q^+)(p^++Q^+)}
\ln{Q^2(p^++Q^+)^2\over m^2Q^+(K^++Q^+)}\bigg\}\\
&\to&{\check\Gamma}_{\Box_L}
-{ig^4N_c\gamma_1^+\gamma_2^+m^2\over 4\pi^2(p^++q^+)Q^+Q^2}
\bigg\{{q^+\over p^+}\sum_{0<K^+<Q^+}{1\over K^+}
\ln{Q^2p^{+2}\over m^2Q^+ K^+}
\nonumber\\
&&-{q^+-Q^+\over p^++Q^+}\sum_{0<K^+<Q^+}{1\over K^+}
\ln{Q^2(p^++Q^+)^2\over m^2Q^+K^+}
\nonumber\\
&&-\int_{-Q^+}^0{dK^+\over p^+}
\ln{Q^2p^{+2}\over m^2Q^+(-K^{+})}
%\nonumber\\&&
-\int_{-Q^+}^0{dK^+\over p^++Q^+}
\ln{Q^2(p^++Q^+)^2\over m^2Q^+(K^++Q^+)}\bigg\}\\
&\to&{\check\Gamma}_{\Box_L}
-{ig^4N_c\gamma_1^+\gamma_2^+m^2\over 4\pi^2(p^++q^+)Q^+Q^2}
\bigg\{{(q^++p^+)Q^+\over p^+(p^++Q^+)}\sum_{0<K^+<Q^+}{1\over K^+}
\ln{Q^2p^+(p^++Q^+)\over m^2Q^+K^{+}}
\nonumber\\&&
+{2q^+p^++(q^+-p^+)Q^+\over p^+(p^++Q^+)}\ln{p^{+}\over p^++Q^{+}}
\sum_{0<K^+<Q^+}{1\over K^+}
\nonumber\\
&&-{Q^+\over p^+}\ln{Q^2p^{+2}\over m^2Q^{+2}}
-{Q^+\over p^++Q^+}\ln{Q^2(p^++Q^+)^2\over m^2Q^{+2}}\bigg\}
\eea
We can combine the left box, half of the box bubble, the left triangle,
wave function and half the gluon  self energy. For the
last three contributions we use (\ref{trianglepluswfdim}), that is we
are including the necessary counterterm
(\ref{counterterm}). Remembering that on shell, $Q^-=-Q^+m^2/[2p^+(p^++Q^+)]$,
we find:
\bea
\Gamma_L&\equiv&\Gamma_{\triangle_L}+{1\over2}M_{\rm SE}+{1\over 2}(
Z_2(p)+Z_2(p+Q)-2){2ig^2\gamma_1^+\gamma_2^+Q^-\over Q^+Q^2}
+\Gamma_{\Box_L}+{1\over2}\Gamma^{\rm Bubble}_\Box\nonumber\\
&\to& {\hat\Gamma}^{\rm Rest}_{\triangle_L}+{\check\Gamma}_{\Box_L}
+{ig^4N_c\gamma_1^+\gamma_2^+Q^-\over4\pi^2Q^2Q^{+}}
\bigg(-{11\over6}\ln\{Q^2\delta e^\gamma\}+{67\over18}
+2\ln{Q^2\over m^2} - 4 +{2\pi^2\over 3}\nonumber\\
&&
+\ln{p^+\over Q^+}-\ln^2{p^+\over Q^+}+\ln{p^++Q^+\over Q^+}
-\ln^2{p^++Q^+\over Q^+}\bigg)
\nonumber\\
&&
-{ig^4N_c\gamma_1^+\gamma_2^+m^2\over 4\pi^2(p^++q^+)Q^+Q^2}
\bigg\{-{Q^+\over p^+}\ln{Q^2p^{+2}\over m^2Q^{+2}}
-{Q^+\over p^++Q^+}\ln{Q^2(p^++Q^+)^2\over m^2Q^{+2}}\bigg\}\nonumber\\
&&
-{ig^4N_c\gamma_1^+\gamma_2^+m^2\over 4\pi^2Q^2p^+(p^++Q^+)}
\left(1+
{2q^+p^++(q^+-p^+)Q^+\over Q^+(p^++q^+)}\ln{p^{+}\over p^++Q^{+}}
\right)\sum_{0<K^+<Q^+}{1\over K^+}\nonumber\\
&\to& {\hat\Gamma}^{\rm Rest}_{\triangle_L}+{\check\Gamma}_{\Box_L}
+{ig^4N_c\gamma_1^+\gamma_2^+Q^-\over4\pi^2Q^2Q^{+}}
\bigg({67\over18}
+2\ln{Q^2\over m^2} - 4 +{2\pi^2\over 3}\nonumber\\
&&
+\ln{p^+\over Q^+}-\ln^2{p^+\over Q^+}+\ln{p^++Q^+\over Q^+}
-\ln^2{p^++Q^+\over Q^+}\bigg)
\nonumber\\
&&
-{ig^4N_c\gamma_1^+\gamma_2^+m^2\over 4\pi^2(p^++q^+)Q^+Q^2}
\bigg\{-{Q^+\over p^+}\ln{Q^2p^{+2}\over m^2Q^{+2}}
-{Q^+\over p^++Q^+}\ln{Q^2(p^++Q^+)^2\over m^2Q^{+2}}\bigg\}\nonumber\\
&&
+\Gamma^{\rm Tree}{g^2N_c\over 4\pi^2}\left(1+
{q^{+2}+p^{+2}\over q^{+2}-p^{+2}}\ln{p^{+}\over q^+}
\right)\sum_{0<K^+<Q^+}{1\over K^+}+\Gamma^{\rm Tree}{g^2N_c\over 8\pi^2}
\bigg(-{11\over6}\ln\{Q^2\delta e^\gamma\}\bigg)
\label{}
\eea
In the final form, we have displayed the ultraviolet and uncanceled infrared
divergences in the last line as a multiple of the tree amplitude.
The amplitude $\Gamma_R$ can be computed directly, but it is more
simply obtained from $\Gamma_L$ through the substitutions $p^+\to
q^+-Q^+, q^+\to p^++Q^+$. But on shell we have $Q^+=q^+-p^+$, so in
in fact $\Gamma_R=\Gamma_L$. Thus
\bea
\Gamma^{\rm 1 Loop}&=&\Gamma_L+\Gamma_R=2\Gamma_L\nonumber\\
&=&\Gamma^{\rm Finite}+\Gamma^{\rm Tree}{g^2N_c\over 4\pi^2}
\bigg\{2\left(1+
{q^{+2}+p^{+2}\over q^{+2}-p^{+2}}\ln{p^{+}\over q^+}
\right)\sum_{0<K^+<Q^+}{1\over K^+}
-{11\over6}\ln\{Q^2\delta e^\gamma\}\bigg\}
\eea
The sign and magnitude of the ultraviolet divergent term agrees exactly
with asymptotic freedom.
We shall see that in its contribution to probabilities the
infrared divergences will be cancelled by contributions from soft
gluon Bremsstrahlung. For this purpose, we need to add the
tree contribution and square the result.
\bea
|A^{\rm Elastic}|^2\sim|\Gamma^{\rm Tree}+\Gamma^{\rm Finite}|^2
\bigg|1-2{g^2N_c\over4\pi^2}
\left({p^{+2}+q^{+2}\over q^{+2}-p^{+2} }\ln{q^+\over p^{+}}
-1\right)\sum_{0<K^+<Q^+}
{1\over K^+}-{11\over6}\ln\{Q^2\delta e^\gamma\}\bigg|^2\nonumber\\
\approx |\Gamma^{\rm Tree}+\Gamma^{\rm Finite}|^2
\bigg[1-2{g^2N_c\over4\pi^2}
\left({p^{+2}+q^{+2}\over q^{+2}-p^{+2} }\ln{q^{+2}\over p^{+2}}
-2\right)\sum_{0<K^+<Q^+}
{1\over K^+}-{11\over3}\ln\{Q^2\delta e^\gamma\}\bigg]
\label{elasticprob}
\eea
\section{Soft Bremsstrahlung and Probabilities}
Soft gluon emission or absorption from scattered branions is dominated by
the diagrams where the emitted or absorbed gluon is directly
attached to external lines. In the context of large $N_c$
we only need sum coherently the two diagrams where the gluon
is attached to neighboring lines, i.e. either emission between
the two outgoing branions or absorption between the incoming
branions. In the first case we simply multiply the amplitude
for the core process by the factor
\bea
&&-g{{\bfs k}\cdot{\bfs\epsilon}\over k^+}\left[{p^++Q^+\over (p+Q)\cdot k}
-{q^+-Q^+\over (q-Q)\cdot k}\right] \noindent\\
&=&2g{{\bfs k}\cdot{\bfs\epsilon}}\left[{(p^++Q^+)^2\over k^{+2}m^2
+{\bfs k}^2(p^++Q^+)^2}
-{(q^+-Q^+)^2\over k^{+2}m^2
+{\bfs k}^2(q^+-Q^+)^2}\right]\equiv
2g{{\bfs k}\cdot{\bfs\epsilon}}\left[{1\over A
+{\bfs k}^2}
-{1\over B
+{\bfs k}^2}\right]
\eea
the relative minus sign arising because the two branions in the
final state have opposite color. Here $A=m^2k^{+2}/(p^++Q^+)^2$
and $B=m^2k^{+2}/(q^+-Q^+)^2$. The probability for gluon emission
is given by squaring the amplitude, summing over color and
gluon spin and integrating over
${\bfs k}, k^+$ in a small window about zero.
\bea
P=|A_{\rm core}|^2\sum_{k^+<k_{\rm max}}\int_{{\bfs k}^2<\Delta^2_T(k^+)} 
d{\bfs k}{4g^2N_c\over8\pi^32k^+}
\left[{{\bfs k}^2\over ({\bfs k}^2+A)^2}+{{\bfs k}^2\over ({\bfs k}^2+B)^2}
-{2{\bfs k}^2\over({\bfs k}^2+A)({\bfs k}^2+B)}\right]
\eea
The integrals are elementary:
\bea
\int_{{\bfs k}^2<\Delta^2_T(k^+)} 
d{\bfs k}{{\bfs k}^2\over({\bfs k}^2+A)({\bfs k}^2+B)}
&=&{\pi\over B-A}\left[B\ln{\Delta_T^2+B\over B}-A\ln{\Delta_T^2+A\over A}
\right]\nonumber\\
\int_{{\bfs k}^2<\Delta^2_T(k^+)} d{\bfs k}{1\over({\bfs k}^2+A)^2}
&=&{\pi}\left[\ln{\Delta_T^2+A\over A}-{\Delta_T^2\over\Delta_T^2+ A}
\right]
\eea
Thus
\bea
P=|A_{\rm core}|^2{g^2N_c\over4\pi^2}\sum_{k^+<k_{\rm max}}{1\over k^+}
\left[{A+B\over A-B}\ln{A(\Delta_T^2+B)\over B(\Delta_T^2+A)}
-{\Delta_T^2\over \Delta_T^2+A}-{\Delta_T^2\over \Delta_T^2+B}\right]
\eea
Next we choose how to specify the resolutions. As discussed
in \cite{chakrabartiqt2} a nice choice is to limit the virtuality
of the two ``jet'' momenta $p+Q+k$ and $q-Q+k$:
\bea
&&\hskip-.5in-(p+Q)\cdot k < \Delta^2, \qquad -(q-Q)\cdot k < \Delta^2 \\
&\rightarrow& {\bfs k}^2<{\rm min}\left(
2k^+{\Delta^2-k^+m^2/2(p^++Q^+)\over(p^++Q^+)}, 
2k^+{\Delta^2-k^+m^2/2(q^+-Q^+)\over(q^+-Q^+)} \right)
\eea 
We could choose the upper limit on $k^+$ independently of $\Delta$
as long as it is less than the least of $2(p^++Q^+)\Delta^2/m^2$,
$2(q^+-Q^+)\Delta^2/m^2$. But for definiteness let's
choose
\beq
k^+<k_{\rm max}\equiv {\rm min}\left\{(q^+-Q^+){\Delta^2\over m^2},
(p^++Q^+){\Delta^2\over m^2}\right\}
\eeq
With resolutions set, we now examine the small $k^+$ limit of the
probability summand. We have required $\Delta_T^2=O(k^+)$ and 
$A,B = O(k^{+2})$, we can neglect $A,B$ in comparison to $\Delta_T$
so we find
\bea
{\rm Summand} \sim {1\over k^+}
\left[{A+B\over A-B}\ln{A\over B}-2\right]=
{1\over k^+}\left[{(p^++Q^+)^2+(q^+-Q^+)^2\over
(p^++Q^+)^2-(q^+-Q^+)^2 }\ln{(p^++Q^+)^2\over(q^+-Q^+)^2 }-2\right]
\eea
Actually, since we are insisting that $Q^+>0$ the on shell condition
is $Q^+=q^+-p^+$ with $q^+>p^+$. And we find the simplification
\bea
P^{\rm Brem}_{\rm IR} \sim 2{g^2N_c\over4\pi^2}|A_{\rm core}|^2
\left[{p^{+2}+q^{+2}\over
q^{+2}-p^{+2}}\ln{q^{+2}\over p^{+2}}-2\right]\sum{1\over k^+}
\eea
where we have added the absorption probability of an extra 
soft gluon in the initial state, which accounts for the factor of 2.
Combining this result with the square of the elastic
amplitude, we see that the infrared divergence cancels.
\section{Concluding Remarks}
In this article we have calculated physical on-shell branion
branion scattering through one loop for the case that the
branions are Dirac fermions living on parallel 1-branes. This
work refines and completes a calculation initiated in \cite{rozowskyt}
by carrying out a careful treatment of the on-shell
limit including a proper definition of scattering probabilities
allowing for the emission and absorption of extra soft
gluons. The ambiguity of the on-shell limit found in \cite{rozowskyt}
came from attempting the continuous $K^+$ limit for an unphysical
off-shell quantity. This is therefore another example of the
novel aspects of lightcone gauge. In a normal covariant gauge
no infrared cutoff is needed when computing off-shell quantities.

We worked on-shell from the beginning in this paper, so the entire calculation
was actually quite different from that carried out  in \cite{rozowskyt}.
Besides this we also used the worldsheet friendly ultraviolet
cutoff of 
\cite{bardakcitmean,bardakcimean,thornscalar,chakrabartiqt1,chakrabartiqt2} 
rather
than the one employed in \cite{rozowskyt}. By comparing our
results to those given by dimensional regularization we were able
to identify all the 1-loop counterterms that will be required
for the construction of the lightcone worldsheet description of
this system. In this concluding section we shall
briefly indicate how the worldsheet formalism can handle
these counterterms. But since 
there remain some unresolved issues in the
worldsheet construction with 1-brane sources, we stress that
it is only illustrative, and the final ``best'' solution may be
quite different.

First of all, the counterterms for the self energy diagrams are
no different than those we required in \cite{chakrabartiqt1,chakrabartiqt2}.
There is of course the branion mass shift (\ref{contsmalldelta}), which 
is nothing but mass renormalization. There is some novelty in 
the fact that the zero thickness
of the 1-brane promotes a single log divergence to a double log one,
but that does not change the fact that the shift is a Lorentz invariant
constant, and mass renormalization proceeds as usual. But there
is also a contribution to the self-energy calculation that is
``tadpole-like'' coming from the instantaneous
longitudinal gluon and that does not involve a propagating intermediate
state. This is just the term we associated with the added semi-circular
contours:
\bea
-i\Sigma^{\rm Instant}&=&g^2N_c\gamma^+\int{d{\bfs K}\over(2\pi)^4}
e^{-\delta({\bfs K}+{\bfs k}_0)^2}
\Bigg[-2i\pi\sum_{0<K^+<p^+}{1\over K^{+2}}\Bigg]\nonumber\\
&=&-{ig^2N_c\gamma^+\over 8\pi^2\delta}\sum_{0<K^+<p^+}{1\over K^{+2}}
=-{ig^2N_c\gamma^+\over 8\pi^2\delta\epsilon}\sum_{n=0}^{M-1}{1\over n^2}
\nonumber\\
&=&-{ig^2N_c\gamma^+\over 8\pi^2\delta\epsilon}\left[
{\pi^2\over6}-{1\over M}+O\left({1\over M^2}\right)\right]
\sim-{ig^2N_c\gamma^+\over 48\delta\epsilon}
+{ig^2N_c\gamma^+\over 8\pi^2\delta p^+}
\eea
The second term has the right behavior to be absorbed in mass renormalization.
The first term is a divergent $p^+$ {\it independent} shift in $p^-$ the
lightcone ``energy'' of the branion. On the lightcone worldsheet it
therefore has the interpretation as a boundary energy or
boundary ``cosmological constant''. Again such a term has
already been encountered in the gluon self-energy
as discussed in \cite{chakrabartiqt1,chakrabartiqt2}, and introduces
no new problems for the lightcone worldsheet.

The branion-gluon vertex counterterm (\ref{counterterm})
looks more problematic because of the nonpolynomial
$p^+$ dependence. It is helpful to rearrange it a little
\bea
\Gamma^{\rm C.T.}&=&{ig^4N_c\gamma_1^+\gamma_2^+\over4\pi^2Q^{+2}}
\Bigg\{-{1\over12}
+\left(1+{2Q^+Q^-\over Q^2}\right)
\bigg[{1\over4Q^+}\sum_{-Q^+<K^+<0}{1\over x(1-x)}-1\bigg]
\Bigg\}
\label{counterterm2}
\eea
Since this expression will be multiplied by $e^{i{\bfs Q}\cdot{\bfs L}}$
and integrated over ${\bfs Q}$, the ${\bfs Q}$ independent terms will
be proportional to $\delta({\bfs L})=0$ for the process we
are analyzing since ${\bfs L}\neq0$. Thus we are left with the
problem of representing 
\bea
{ig^4N_c\gamma_1^+\gamma_2^+Q^-\over4\pi^2Q^{+}Q^2}
\bigg[\sum_{0<k^+<Q^+}{1\over k^+}
-2\bigg]
\eea
locally on the worldsheet. Although awkward looking, there is a way to do
it. First of all a factor of $1/k^+$ can be produced by the insertion
of a local worldsheet field, call it $\phi(\sigma,\tau)$ 
at a point a distance $k^+$ from the
boundary of the strip representing the gluon propagator (see 
\cite{bardakcit} in connection with the representation
of $1/p^+$ factors in vertex functions). Then integrating
this point across the gluon strip reproduces the desired nonpolynomial
terms. A truly local prescription, however, should integrate the
field insertion point over the whole worldsheet, not just
a single time slice on a single propagator. So we need to arrange
things so that the integral over the whole worldsheet contributes
only at one time and only on the gluon propagator. Again there is
precedent for this sort of effect in the way the worldsheet can
produce quartic vertices. Briefly the way this works is that
one can introduce freely 
any number of extra worldsheet fields ${\bfs \chi}_i$ which satisfy
${\bfs \chi}_i=0$ on all boundaries together with ghost fields
$\beta_i,\gamma_i$ such that the path integral over them all gives
unity. Denoting $\partial/\partial\sigma$ by
${}^\prime$, then $\VEV{{\bfs \chi}^\prime_i}=0$ but 
$\VEV{{\bfs \chi}^\prime_i(\sigma,\tau)
{\bfs \chi}^\prime_i(\sigma^\prime,\tau^\prime)}
\propto\delta(\tau-\tau^\prime)$.
By attaching one of these extra fields to the branion gluon 
interaction point and another to the local field $\phi$, the
contribution can be restricted in the desired way. We content ourselves
here with this feasibility argument and leave a definitive
solution for future work, in which we hope to resolve  the
other difficulties posed by the introduction
of 1-brane sources into the lightcone worldsheet formalism.
\vskip14pt
\noindent\underline{ Acknowledgments}: 
I would like to thank Jian Qiu for valuable discussions. 
This research was supported in part by the Department
of Energy under Grant No. DE-FG02-97ER-41029.

\appendix
\section{Triangle Integrals without Numerator Factors}
In lightcone evaluations we always reserve the $K^+$ integrations till
last. Starting with the left triangle integrand,
we do the $K^-$ integral first:
\bea
\int {dK^-\over 2\pi}{1\over K^2(K+Q)^2(m^2+(p-K_\|)^2}
&=&{i\theta(p^+-K^+)\theta(K^+)\over2(p^+-K^+)[{\bfs K}^2+
B][({\bfs K}+{\bfs Q})^2+C]}\nonumber\\
&&+{i\theta(Q^++K^+)\theta(-K^+)(K^++Q^+)\over2Q^+(p^+-K^+)
[({\bfs K}^\prime-{\bfs Q}^\prime)^2+D][({\bfs K}^\prime)^2+C]}\\
B={m^2K^{+2}\over p^+(p^+-K^+)}\qquad C&=&{m^2(K^++Q^+)^2\over(p^++Q^+)
(p^+-K^+)}\nonumber\\
 \hskip-.25in D=-K^+(K^++Q^+)\left[{{\bfs Q}^2\over Q^{+2}}
+{m^2\over p^+(p^++Q^+)}\right]&=&-{K^+(K^++Q^+)\over
Q^{+2}}Q^2\nonumber\\ 
{\bfs K}^\prime={\bfs K}+{\bfs Q} &&
\qquad{\bfs Q}^\prime={Q^++K^+\over Q^+}{\bfs Q}
\eea
Next the transverse momentum integral can be done after combining denominators
with the Feynman trick:
\bea
&&\hskip-.4in\int{d^2{\bfs K}\over 4\pi^2}\int_0^1dx{1\over[({\bfs K}+x{\bfs Q})^2
+x(1-x){\bfs Q}^2+B(1-x)+Cx]^2}={1\over4\pi}\int_0^1dx
{1\over x(1-x){\bfs Q}^2+B(1-x)+Cx}\nonumber\\
&&\hskip-.4in\int{d^2{\bfs K}\over 4\pi^2}\int_0^1dx
{1\over[({\bfs K}^\prime-x{\bfs Q}^\prime)^2
+x(1-x){\bfs Q}^{\prime2}+Dx+C(1-x)]^2}={1\over4\pi}\int_0^1dx
{1\over x(1-x){\bfs Q}^{\prime2}+Dx+C(1-x)}\nonumber
\label{transverseint}
\eea
The $x$ integral can be done by factoring the denominator:
\bea
\int_0^1 dx {1\over ax(1-x) +bx +c(1-x)}&=&{1\over a(r_+-r_-)}
\ln{r_+(1-r_-)\over-r_-(r_+-1)}\\
r_{\pm}&=&{1\over2}+{b-c\over2a}\pm{1\over2a}
\sqrt{a^2+b^2+c^2+2a(b+c)-2bc}
\label{xint}
\eea
Let us denote the roots for $a={\bfs Q}^2$ and $b=B$, $c=C$
by $r_\pm$ without primes, and the roots with $a={\bfs Q}^{\prime2}$
and $b=C$, $c=D$ by $r^\prime_\pm$.
Then
\bea
I_L&\equiv&\int {d^2{\bfs K}dK^-\over 8\pi^3}{1\over K^2(K+Q)^2(m^2+(p-K_\|)^2}
\nonumber\\
&=&{1\over8\pi(p^+-K^+)}\bigg[
{i\theta(p^+-K^+)\theta(K^+)\over {\bfs Q}^2(r_+-r_-)}
\ln{r_+(1-r_-)\over-r_-(r_+-1)}\nonumber\\
&&\qquad +{i\theta(Q^++K^+)\theta(-K^+)(K^++Q^+)\over Q^+{\bfs Q}^{\prime2}
(r^\prime_+-r^\prime_-)}
\ln{r^\prime_+(1-r^\prime_-)\over-r^\prime_-(r^\prime_+-1)}\bigg]\\
&\equiv&\theta(p^+-K^+)\theta(K^+)I^1_L+\theta(Q^++K^+)\theta(-K^+)I^2_L\\
I^1_L&=&{i\over8\pi(p^+-K^+){\bfs Q}^2(r_+-r_-)}
\ln{r_+(1-r_-)\over-r_-(r_+-1)}\\
I^2_L&=&{iQ^+\over8\pi(p^+-K^+)(K^++Q^+){\bfs Q}^2(r^\prime_+-r^\prime_-)}
\ln{r^\prime_+(1-r^\prime_-)\over-r^\prime_-(r^\prime_+-1)}
\eea
The $K^+$ integration of these results is infrared divergent for
$K^+$ near 0 and $-Q^+$. The singular behavior for $K^+$ near
$p^+$ is integrable. To extract the infrared structure we
examine the behavior of $I_L$ near each of these dangerous points.

Consider first $K^+\sim0$ from the positive side. Then 
$B\sim0$ and $r_+\to 1$, $r_+-1\sim B/({\bfs Q}^2+C)$, $r_-\to-C/{\bfs Q}^2$,
$1-r_-\to ({\bfs Q}^2+C)/{\bfs Q}^2$, ${\bfs Q}^2+C\to {\bfs Q}^2+m^2Q^{+2}/p^+(p^++Q^+)
={\bf Q}^2+Q_\|^2=Q^2$. and
\bea
I_L\sim {i\over8\pi p^+ Q^2}\ln{Q^4p^{+2}\over m^2Q_\|^2K^{+2}}\qquad 
{\rm for}~ K^+\to 0_+
\eea
It is simple to check that $I_L=O(1)$ as $K^+\to p^+$.
Next we consider $K^+\sim0$ from below. Then $D\sim0$, $r_-^\prime\to0$,
$r_+^\prime\to(A+C)/A$, and
\bea
I_L\sim {i\over8\pi p^+ Q^2}\ln{Q^2Q^{+}\over Q_\|^2(-K^{+})}\qquad 
{\rm for}~ K^+\to 0_-
\eea
Finally, we consider $K^+\sim -Q^+$. In this case, $D\sim Q^2(K^++Q^+)/Q^+$,
$C\sim m^2(K^++Q^+)^2/(p^++Q^+)^2$, so $r_+^\prime-1\sim C/D$,
$r_-^\prime\sim -D/{\bfs Q}^{\prime2}$, so
\bea
I_L\sim{i\over8\pi (p^++Q^+) Q^2}
\ln{Q^2(p^++Q^+)^2\over m^2Q^+(K^++Q^+)}\qquad {\rm for}~ K^+\to -Q^+
\eea

Turning now to the right triangle integrand,
we do the $K^-$ integral first:
\bea
\int {dK^-\over 2\pi}{1\over K^2(K+Q)^2(m^2+(q+K_\|)^2}
&=&{i\theta(q^++K^+)\theta(-Q^+-K^+)\over2(q^++K^+)[{\bfs K}^2+
{\bar B}][({\bfs K}+{\bfs Q})^2+{\bar C}]}\nonumber\\
&&+{i\theta(Q^++K^+)\theta(-K^+)(-K^+)\over2Q^+(q^++K^+)
[({\bfs K}+{\bfs Q}^\prime)^2+{\bar D}][{\bfs K}^2+{\bar B}]}\\
{\bar B}={m^2K^{+2}\over q^+(q^++K^+)}\qquad {\bar C}
&=&{m^2(K^++Q^+)^2\over(q^+-Q^+)(q^++K^+)}\nonumber\\
 \hskip-.25in {\bar D}=-K^+(K^++Q^+)\left[{{\bfs Q}^2\over Q^{+2}}
+{m^2\over q^+(q^+-Q^+)}\right]&=&-{K^+(K^++Q^+)\over Q^{+2}}Q^2 = D
\qquad{\bfs Q}^\prime=-{K^+\over Q^+}{\bfs Q}
\eea
Next the transverse momentum integrals are given by (\ref{transverseint})
with appropriate substitutions, and
the $x$ integral by (\ref{xint}).
Let us denote the roots for $a={\bfs Q}^2$ and $b={\bar B}$, $c={\bar C}$
by ${\bar r}_\pm$ without primes, and the roots with $a={\bfs Q}^{\prime2}$
and $b={\bar C}$, $c={\bar D}$ by ${\bar r}^\prime_\pm$.
Then
\bea
I_R&\equiv&\int {d^2{\bfs K}dK^-\over 8\pi^3}{1\over K^2(K+Q)^2
(m^2+(q+K_\|)^2)}
\nonumber\\
&=&{1\over8\pi(q^++K^+)}\bigg[
{i\theta(q^++K^+)\theta(-K^+-Q^+)\over {\bfs Q}^2({\bar r}_+-{\bar r}_-)}
\ln{{\bar r}_+(1-{\bar r}_-)\over-{\bar r}_-({\bar r}_+-1)}\nonumber\\
&&\qquad +{i\theta(Q^++K^+)\theta(-K^+)(-K^+)\over Q^+{\bfs Q}^{\prime2}
({\bar r}^\prime_+-{\bar r}^\prime_-)}
\ln{{\bar r}^\prime_+(1-{\bar r}^\prime_-)\over-{\bar r}^\prime_-
({\bar r}^\prime_+-1)}\bigg]\\
&\equiv&\theta(q^++K^+)\theta(-K^+-Q^+)I^1_R
+\theta(Q^++K^+)\theta(-K^+)I^2_R\\
I^1_R&=&{i\over8\pi(q^++K^+){\bfs Q}^2({\bar r}_+-{\bar r}_-)}
\ln{{\bar r}_+(1-{\bar r}_-)\over-{\bar r}_-({\bar r}_+-1)}
\qquad\quad -q^+<K^+<-Q^+\\
I^2_R&=&{iQ^+\over8\pi(q^++K^+)(-K^+){\bfs Q}^2
({\bar r}^\prime_+-{\bar r}^\prime_-)}\ln{{\bar r}^\prime_+(1-{\bar r}^\prime_-)\over-{\bar r}^\prime_-({\bar r}^\prime_+-1)}
\qquad -Q^+<K^+<0
\eea
Again the $K^+$ integration is infrared divergent for
$K^+$ near 0 and $-Q^+$. The singular behavior for $K^+$ near
$-q^+$ is integrable. To extract the infrared structure we
examine the behavior of $I_R$ near each of these dangerous points.

Consider first $K^+\sim -Q^+$ from the negative side. Then 
${\bar C}\sim0$ and ${\bar r}_+\to ({\bfs Q}^2+{\bar B})/{\bfs Q}^2$, 
${\bar r}_+-1
\sim {\bar B}/{\bfs Q}^2$, ${\bar r}_-\to-{\bar C}/({\bfs Q}^2+{\bar B})$,
$1-r_-\to 1$, ${\bfs Q}^2+{\bar B}\to Q^2$. and
\bea
I_R\sim {i\over8\pi (q^+-Q^+) Q^2}\ln{Q^4(q^{+}-Q^+)^2\over m^2Q_\|^2
(K^{+}+Q^+)^2}\qquad 
{\rm for}~ K^+ +Q^+\to 0_-
\eea
It is simple to check that $I_R=O(1)$ as $K^+\to -q^+$.
Next we consider $K^+\sim -Q^+$ from above. Then ${\bar D}\sim0$, 
$r_-^\prime\to0$,
$r_+^\prime\to({\bfs Q}^2+{\bar B})/{\bfs Q}^2$, and
\bea
I_R\sim {i\over8\pi (q^+-Q^+) Q^2}\ln{Q^2Q^{+}\over Q_\|^2(K^{+}+Q^+)}\qquad 
{\rm for}~ K^++Q^+\to0_+
\eea
Finally, we consider $K^+\sim 0_-$. In this case, ${\bar D}\sim -Q^2K^+/Q^+$,
${\bar B}\sim m^2K^{+2}/q^{+2}$, so ${\bar r}_+^\prime-1\sim 
{\bar B}/{\bar D}$, $r_-^\prime\sim -{\bar D}/{\bfs Q}^{\prime2}$, so
\bea
I_R\sim{i\over8\pi q^+Q^2}
\ln{Q^2q^{+2}\over m^2Q^+(-K^+)}\qquad {\rm for}~ K^+\to 0_-
\eea
We have remarked in the text that right triangle integrals can be
obtained from left triangle integrals through the substitutions
$p\to q-Q$, $q\to p+Q$. In the context of the integrals in this section
which have left $K^+$ integration unperformed, we see by
direct inspection that $I^{1,2}_R(q^+,K^+)=I^{1,2}_L(q^+-Q^+,-K^+-Q^+)$.
Note in this context that the range $-q^+<K^+<0$ can be expressed
as $-Q^+<-K^+-Q^+<q^+-Q^+$, analogous to the range $-Q^+<K^+<p^+$.
\section{Triangle Integrals with Numerator Factors}
Some of the triangle integrals we need contain numerator factors
involving $K^-$, ${\bfs K}^2$ or ${\bfs K}\cdot{\bfs Q}$. In the
text we have shown how to replace $K^-$ factors with polynomials
in $K^+$ together with cancelled propagator terms, which have
the structure of bubble diagrams, explicitly evaluated in the
text. Once the numerators are free of $K^-$ factors, the $K^-$ integration
is done by contours, leaving denominators which are quadratics
in ${\bfs K}$. We can then replace numerator factors of 
${\bfs K}^2$ or ${\bfs K}\cdot{\bfs Q}$ with functions of $K^+$ times the
integrals of the previous section plus integrals with only one 
denominator. We evaluate these one denominator integrals
in this section. Since they are log divergent in the UV we give 
both delta regulator and dimensional regulator form of the answers. 

After integration over $K^-$ there are four distinct transverse integrals
to do: the left and right triangle integrals and for each of these,
to distinct regions of $K^+$. For each of these four transverse integrals
there can be three numerators; $1$, ${\bfs K}^2$, ${\bfs K}\cdot{\bfs Q}$.
We adopt the notation $I^{1,2}_{L,R}[X]$ with $X$ symbolizing the
numerator. In the previous section we evaluated all of the
$I^{1,2}_{L,R}[1]\equiv I^{1,2}_{L,R}$.

The one denominator integrals have the general form
\bea
\int {d^2 K\over(2\pi)^2}{1\over ({\bfs K}+{\bfs L})^2 + Z}&\to&
\int {d^2 K\over(2\pi)^2}
{e^{-\delta({\bfs K}+{\bfs k}_0)^2}\over ({\bfs K}+{\bfs L})^2 + Z}\qquad 
\delta-{\rm reg}\\
&\to&\int {d^d K\over(2\pi)^d}{1\over ({\bfs K}+{\bfs L})^2 + Z}\qquad
{\rm dim-reg}
\eea
with $\delta$ and dimensional regularization respectively. In the
first case we have
\bea
\int {d^2 K\over(2\pi)^2}
{e^{-\delta({\bfs K}+{\bfs k}_0)^2}\over ({\bfs K}+{\bfs L})^2 + Z}
&=&\int_0^\infty dT\int {d^2 K\over(2\pi)^2}e^{-\delta({\bfs K}+{\bfs k}_0)^2
-T[({\bfs K}+{\bfs L})^2 + Z]}\nonumber\\
&=&{1\over4\pi}\int_0^\infty {dT\over T+\delta} 
\exp\left\{-TZ-{T\delta\over T+\delta}({\bfs L}-{\bfs k}_0)^2\right\}
\eea
The second term in the exponent is $O(\delta)$ for all $T$ and the
divergence as $\delta\to0$ is only logarithmic, so this term is negligible
for $\delta\sim0$. In the limit these integrals are therefore independent
of ${\bfs L}$ and ${\bfs k}_0$.
\bea
\int {d^2 K\over(2\pi)^2}
{e^{-\delta({\bfs K}+{\bfs k}_0)^2}\over ({\bfs K}+{\bfs L})^2 + Z}
&\sim&{1\over4\pi}\int_0^\infty {dT\over T+\delta} 
e^{-TZ}\ \sim\ -{1\over 4\pi}\ln(Z\delta e^\gamma)
\eea
In dim-reg, we have simply
\bea
\int {d^d K\over(2\pi)^d}{1\over ({\bfs K}+{\bfs L})^2 + Z}
={\Gamma(1-d/2)\over(4\pi)^{d/2}Z^{(d-2)/2}}
\sim {1\over 4\pi}\left[{\Gamma(1-d/2)\over(4\pi)^{(d-2)/2}}
-\ln Z\right]\to -{1\over 4\pi}\ln(Z\delta e^\gamma)
\eea
with the correspondence (\ref{dimdelta}). We see that the two
regularizations exactly agree for these integrals.

It remains to obtain the eight distinct integrals with nontrivial numerators.
\bea
I^1_L[{\bfs K}^2]&=&-BI^1_L + I^1_L[{\bfs K}^2+B]
=-BI^1_L-{i\over8\pi(p^+-K^+)}\ln(C\delta e^\gamma)
\label{ksquared1}\\
I^1_L[{\bfs K}\cdot{\bfs Q}]&=&{1\over2}I^1_L[({\bfs K}+{\bfs Q})^2+C
-{\bfs K}^2-B]
+{1\over2}(B-C-{\bfs Q}^2)I^1_L\nonumber\\
&=&{1\over2}(B-C-{\bfs Q}^2)I^1_L-{i\over16\pi(p^+-K^+)}\ln(B/C)\\
I^2_L[{\bfs K}^2]&=&I^2_L[{\bfs K}^{\prime2}-2{\bfs K}^\prime\cdot{\bfs Q}]
+{\bfs Q}^2I^2_L\nonumber\\
&=&-\left({K^+C\over K^++Q^+}+{K^+\over Q^+}{\bfs Q}^2+{Q^+D\over K^++Q^+}
\right)I^2_L\nonumber\\
&&-{i\over8\pi (p^+-K^+)}\left({K^++Q^+\over Q^+}
\ln(D\delta e^\gamma)+\ln{C\over D}\right)\label{ksquared2}\\
I^2_L[{\bfs K}\cdot{\bfs Q}]&=&\left({\bfs Q}^2{K^+-Q^+\over2Q^+}-
{Q^+(C-D)\over2(K^++Q^+)}\right)I^2_L+{i\over16\pi(p^+-K^+)}\ln{C\over D}\\
I^1_R[{\bfs K}^2]&=&-{\bar B}I^1_R-{i\over 8\pi(q^++K^+)}
\ln({\bar C}\delta e^\gamma)\\
I^1_R[{\bfs K}\cdot{\bfs Q}]&=&{1\over2}({\bar B}-{\bar C}-{\bfs Q}^2)I^1_R
-{i\over16\pi(q^++K^+)}\ln({\bar B}/{\bar C})\\
I^2_R[{\bfs K}^2]&=&-{\bar B}I^2_R+{iK^+\over8\pi Q^+(q^++K^+)}
\ln(D\delta e^\gamma)\\
I^2_R[{\bfs K}\cdot{\bfs Q}]&=&-{Q^+\over2K^+}
\left({\bar B}-D-{K^{+2}\over Q^{+2}}
{\bfs Q}^2\right)I^2_R-{i\over16\pi(q^++K^+)}\ln{{\bar B}\over D}
\eea
These results are used in the evaluation of not only the
triangle diagrams themselves, but also triangle-like integrals
that contribute to the box diagrams. They must still be integrated
over $K^+$, for which there are several potential divergences when
$K^+\to p^+,0,-Q^+,-q^+$. The behavior at these points is
tabulated in Figs.~\ref{lasympt}, \ref{rasympt}. The points $p^+$
and $-q^+$ cause no difficulty. This is because 
${\bfs K}^2$ is always multiplied by $p^+-K^+$ or $q^++K^+$
whenever it occurs in the left or right triangle diagram respectively.
The points $0$, and $-Q^+$ can cause infrared divergences. However,
the integrals contributing to the actual triangle diagrams
turn out to be convergent. As we discuss in the text,
there are some residual infrared divergences in the triangle-like
integrals contributing to the box diagram. Inspection of theses
tables easily allows their extraction.

Finally, we have remarked several times in the text 
that right triangle integrals can be
obtained from left triangle integrals through the substitutions
$p\to q-Q$, $q\to p+Q$. In the context of the integrals in these
appendices, which have left $K^+$ integration unperformed, we see by
direct inspection that $I^{1,2}_R(q^+,K^+)=I^{1,2}_L(q^+-Q^+,-K^+-Q^+)$.
Note in this context that the range $-q^+<K^+<0$ can be expressed
as $-Q^+<-K^+-Q^+<q^+-Q^+$, analogous to the range $-Q^+<K^+<p^+$.
There are similar relations between the other integrals. All
together we have 
\bea
I^{1,2}_R(q^+,K^+)
&=&I^{1,2}_L(q^+-Q^+,-K^+-Q^+)\\
I^{1,2}_R[({\bfs K}+{\bfs Q})^2](q^+,K^+)
&=&I^{1,2}_L[{\bfs K}^2](q^+-Q^+,-K^+-Q^+)\\
I^{1,2}_R[-({\bfs K}+{\bfs Q})\cdot{\bfs Q}](q^+,K^+)
&=& I^{1,2}_L[{\bfs K}\cdot{\bfs Q}](q^+-Q^+,-K^+-Q^+)
\eea
After the $K^+$ integrals have been performed, these relations
just produce the general substitution rule quoted above.
\begin{figure}
\begin{center}
\begin{tabular}{|c|c|c|c|}
\hline
\multicolumn{4}{|c|}{{\bf Left Triangle Asymptotics}, $-Q^+<K^+<p^+$} \\
\hline
& & & \\[-.25cm]
& $K^+\to p^+$ & $K^+\to 0$ & $K^+\to -Q^+$ \\[.25cm]
\hline
& & & \\[-.25cm]
$I^1_L$
&$
%\displaystyle 
{i\over8\pi m^2Q^+}\ln{p^++Q^+\over p^+}$  &$
%\displaystyle
{i\over8\pi p^+ Q^2}\ln{Q^4p^{+2}\over m^2Q_\|^2K^{+2}}$ & NA  \\[.25cm]
\hline
& & & \\[-.25cm]
$I^1_L[{\bfs K}^2]$
&$-{i\over8\pi(p^+-K^+)}\bigg[{p^+\over Q^+}\ln{p^++Q^+\over p^+} $
 &$-{i\over8\pi p^+}\ln(Q_\|^2\delta e^\gamma)$ & NA \\[.25cm]
& & &\\[-.25cm]
&$+\ln{(p^++Q^+)m^2\delta e^\gamma\over p^+-K^+}\bigg] $ & &  \\[.25cm]
\hline
& & &\\[-.25cm]
$I^1_L[{\bfs K}\cdot{\bfs Q}]$
&$-{i{\bfs Q}^2\over16\pi Q^+}\ln{p^++Q^+\over p^+}$ &
$-{i\over16\pi p^+}\ln{Q^4\over Q_\|^4}$ & NA\\[.25cm]
\hline
& & &\\[-.25cm]
$I^2_L$
&NA  &$
%\displaystyle
{i\over8\pi p^+ Q^2}\ln{Q^2Q^{+}\over Q_\|^2(-K^{+})}$ & 
$
%\displaystyle 
{i\over8\pi (p^++Q^+) Q^2}
\ln{Q^2(p^++Q^+)^2\over m^2Q^+(K^++Q^+)}$\\[.25cm]
\hline
& & &\\[-.25cm]
$I^2_L[{\bfs K}^2]$
& NA &$-{i\over8\pi p^+}\ln(Q_\|^2\delta e^\gamma)$ &
$
%\displaystyle 
{i{\bfs Q}^2\over8\pi (p^++Q^+) Q^2}
\ln{Q^2(p^++Q^+)^2\over m^2Q^+(K^++Q^+)}$  \\[.25cm]
\hline
& & &\\[-.25cm]
$I^2_L[{\bfs K}\cdot{\bfs Q}]$
& NA &$-{i\over16\pi p^+}\ln{Q^4\over Q_\|^4}$ 
&$
%\displaystyle 
{-i{\bfs Q}^2\over16\pi (p^++Q^+) Q^2}
\ln{Q^2(p^++Q^+)^2\over m^2Q^+(K^++Q^+)}$  \\[.25cm]
\hline
\end{tabular}
\end{center}
\caption{Asymptotic behavior of the left triangle $K^+$ integrands
near singular points.} 
\label{lasympt}
\end{figure}

\begin{figure}
\begin{center}
\begin{tabular}{|c|c|c|c|}
\hline
\multicolumn{4}{|c|}{{\bf Right Triangle Asymptotics}, $-q^+<K^+<0$} \\
\hline
& & & \\[-.25cm]
& $K^+\to -q^+$ & $K^+\to -Q^+$ & $K^+\to 0$ \\[.25cm]
\hline
& & & \\[-.25cm]
$I^1_R$
&$
%\displaystyle 
{i\over8\pi m^2Q^+}\ln{q^+\over q^+-Q^+}$  &$
%\displaystyle
{i\over8\pi (q^+-Q^+) Q^2}\ln{Q^4(q^{+}-Q^+)^2
\over m^2Q_\|^2K^{+2}}$ & NA  \\[.25cm]
\hline
& & & \\[-.25cm]
$I^1_R[{\bfs K}^2]$
&$-{i\over8\pi(q^++K^+)}\bigg[{q^+\over Q^+}\ln{q^+\over q^+-Q^+} $
 &${\bfs Q}^2I^1_R
-{i\over8\pi (q^+-Q^+)}\ln{Q^4\delta e^\gamma\over Q_\|^2}$ & NA \\[.25cm]
& & &\\[-.25cm]
&$+\ln{(q^+-Q^+)m^2\delta e^\gamma\over q^++K^+}\bigg] $ & &  \\[.25cm]
\hline
& & &\\[-.25cm]
$I^1_R[{\bfs K}\cdot{\bfs Q}]$
&$-{i{\bfs Q}^2\over16\pi Q^+}\ln{q^+\over q^+-Q^+}$ &
$-{\bfs Q}^2I^1_R+{i\over16\pi (q^+-Q^+)}\ln{Q^4\over Q_\|^4}$ & NA\\[.25cm]
\hline
& & &\\[-.25cm]
$I^2_R$
&NA  &$
%\displaystyle
{i\over8\pi (q^+-Q^+) Q^2}\ln{Q^2Q^{+}\over Q_\|^2(K^{+}+Q^+)}$ & 
$
%\displaystyle 
{i\over8\pi q^+ Q^2}
\ln{Q^2q^{+2}\over m^2Q^+(-K^+)}$\\[.25cm]
\hline
& & &\\[-.25cm]
$I^2_R[{\bfs K}^2]$
& NA & ${\bfs Q}^2I^2_R
-{i\over8\pi (q^+-Q^+)}\ln{Q^4\delta e^\gamma\over Q_\|^2}$&
$0$  \\[.25cm]
\hline
& & &\\[-.25cm]
$I^2_R[{\bfs K}\cdot{\bfs Q}]$
& NA &$-{\bfs Q}^2I^2_R+{i\over16\pi(q^+-Q^+)}\ln{Q^4\over Q_\|^4}$ 
&$0$  \\[.25cm]
\hline
\end{tabular}
\end{center}
\caption{Asymptotic behavior of the right triangle $K^+$ integrands
near singular points.} 
\label{rasympt}
\end{figure}


\newpage
\begin{thebibliography}{1}
\bibitem{rozowskyt}
  J.~S.~Rozowsky and C.~B.~Thorn,
  %``Defining the force between separated sources on a light front,''
  Phys.\ Rev.\ D {\bf 60} (1999) 045001
  [arXiv:hep-th/9902145].
\bibitem{thooftlargen}
G. 't Hooft, {\sl Nucl. Phys.} {\bf B72} (1974) 461.
\bibitem{klebanovmt}
I.~R.~Klebanov, J.~Maldacena and C.~B.~Thorn,
%  ``Dynamics of Flux Tubes in Large N Gauge Theories,''
JHEP {\bf 0604} (2006) 024
  [arXiv:hep-th/0602255].
  %%CITATION = JHEPA,0604,024;%%
\bibitem{maldacena}
J.~Maldacena,
%``The large-N limit of superconformal field theories and supergravity,''
Adv.\ Theor.\ Math.\ Phys.\ {\bf 2}, 231 (1998),
hep-th/9711200;\\
%%CITATION = HEP-TH 9711200;%%
S.S.~Gubser, I.R.~Klebanov and A.M.~Polyakov,
%``Gauge theory correlators from non-critical string theory,''
Phys.\ Lett.\ {\bf B428}, 105 (1998),
hep-th/9802109;\\
%%CITATION = HEP-TH 9802109;%%
E.~Witten,
%``Anti-de Sitter space and holography,''
Adv.\ Theor.\ Math.\ Phys.\ {\bf 2}, 253 (1998),
hep-th/9802150.
%%CITATION = HEP-TH 9802150;%%
\bibitem{maldacenaqqbar}
S.-J. Rey and J. Yee, 
%{``Macroscopic Strings as Heavy Quarks of
%Large N Gauge Theory and Anti-de Sitter Supergravity,''}
{hep-th/9803001};\\ 
J. Maldacena, 
%{``Wilson loops in large N field
%theories,''} 
Phys. Rev. Lett. {\bf 80}, 4859 (1998),
{hep-th/9803002}.
\bibitem{callang}
C.~G.~Callan and A.~Guijosa,
%``Undulating strings and gauge theory waves,''
Nucl.\ Phys.\ B {\bf 565} (2000) 157
[arXiv:hep-th/9906153].
%%CITATION = HEP-TH 9906153;%%
\bibitem{ericksonsz}
J.~K.~Erickson, G.~W.~Semenoff and K.~Zarembo,
%``Wilson loops in N = 4 supersymmetric Yang-Mills theory,''
Nucl.\ Phys.\ B {\bf 582} (2000) 155
[arXiv:hep-th/0003055].
\bibitem{grossd}
N.~Drukker and D.~J.~Gross,
%  ``An exact prediction of N = 4 SUSYM theory for string theory,''
  J.\ Math.\ Phys.\  {\bf 42} (2001) 2896
  [arXiv:hep-th/0010274].
  %%CITATION = HEP-TH 0010274;%%
\bibitem{ericksonssz}
J.~K.~Erickson, G.~W.~Semenoff, R.~J.~Szabo and K.~Zarembo,
%``Static potential in N = 4 supersymmetric Yang-Mills theory,''
Phys.\ Rev.\ D {\bf 61} (2000) 105006
[arXiv:hep-th/9911088].
\bibitem{browertt}
R. Brower, C-I Tan, and C. B. Thorn,
 %``String / flux tube duality on the lightcone,''
  Phys.\ Rev.\  D {\bf 73} (2006) 124037
  [arXiv:hep-th/0603256].
  %%CITATION = PHRVA,D73,124037;%%
\bibitem{bardakcit}
  K.~Bardakci and C.~B.~Thorn,
  %``A worldsheet description of large N(c) quantum field theory,''
  Nucl.\ Phys.\ B {\bf 626} (2002) 287
  [arXiv:hep-th/0110301].
  %%CITATION = HEP-TH 0110301;%%
\bibitem{thorngauge}
  C.~B.~Thorn,
  %``A worldsheet description of planar Yang-Mills theory,''
  Nucl.\ Phys.\ B {\bf 637} (2002) 272
  [Erratum-ibid.\ B {\bf 648} (2003) 457]
  [arXiv:hep-th/0203167].
  %%CITATION = HEP-TH 0203167;%%
\bibitem{gudmundssontt}
  S.~Gudmundsson, C.~B.~Thorn and T.~A.~Tran,
  %``BT worldsheet for supersymmetric gauge theories,''
  Nucl.\ Phys.\ B {\bf 649} (2003) 3
  [arXiv:hep-th/0209102].
  %%CITATION = HEP-TH 0209102;%%
\bibitem{bardakcitmean}
  K.~Bardakci and C.~B.~Thorn,
  %``A mean field approximation to the world sheet 
%model of planar phi**3  field
  %theory,''
  Nucl.\ Phys.\  B {\bf 652} (2003) 196
  [arXiv:hep-th/0206205];
  %``An improved mean field approximation on the worldsheet for planar phi**3
  %theory,''
  Nucl.\ Phys.\  B {\bf 661} (2003) 235
  [arXiv:hep-th/0212254].
  %%CITATION = NUPHA,B652,196;%%
\bibitem{bardakcimean}
  K.~Bardakci,
  %``Self consistent field method for planar phi**3 theory,''
  Nucl.\ Phys.\  B {\bf 677} (2004) 354
  [arXiv:hep-th/0308197]; 
  %``Mean field approximation for field theories on the worldsheet revisited,''
  Nucl.\ Phys.\  B {\bf 698} (2004) 202
  [arXiv:hep-th/0404076];
  %``Field theory on the world sheet: Mean field expansion and cutoff
  %dependence,''
  arXiv:hep-th/0701098.
\bibitem{trant}
  C.~B.~Thorn and T.~A.~Tran,
  %``The fishnet as anti-ferromagnetic phase of worldsheet Ising spins,''
  Nucl.\ Phys.\  B {\bf 677} (2004) 289
  [arXiv:hep-th/0307203].
  %%CITATION = NUPHA,B677,289;%%
\bibitem{metsaevtt}
  R.~R.~Metsaev, C.~B.~Thorn and A.~A.~Tseytlin,
  %``lightcone superstring in AdS space-time,''
  Nucl.\ Phys.\ B {\bf 596} (2001) 151
  [arXiv:hep-th/0009171].
  %%CITATION = HEP-TH 0009171;%%
%\cite{Goddard:1973qh}
\bibitem{goddardgrt}
  P.~Goddard, J.~Goldstone, C.~Rebbi and C.~B.~Thorn,
  %``Quantum dynamics of a massless relativistic string,''
  Nucl.\ Phys.\  B {\bf 56} (1973) 109.
  %%CITATION = NUPHA,B56,109;%%
\bibitem{thornscalar}
  C.~B.~Thorn,
  %``Renormalization of quantum fields on the lightcone worldsheet. I: Scalar
  %fields,''
  Nucl.\ Phys.\ B {\bf 699} (2004) 427
  [arXiv:hep-th/0405018].
  %%CITATION = HEP-TH 0405018;%%
%\cite{chakrabarti:2005ny}
\bibitem{chakrabartiqt1}
  D.~Chakrabarti, J.~Qiu and C.~B.~Thorn,
  %``Scattering of glue by glue on the light-cone worldsheet. I: Helicity
  %non-conserving amplitudes,''
  Phys.\ Rev.\  D {\bf 72} (2005) 065022
  [arXiv:hep-th/0507280].
  %%CITATION = PHRVA,D72,065022;%%
%\cite{Chakrabarti:2006mb}
\bibitem{chakrabartiqt2}
  D.~Chakrabarti, J.~Qiu and C.~B.~Thorn,
  %``Scattering of glue by glue on the light-cone worldsheet. II: Helicity
  %conserving amplitudes,''
  Phys.\ Rev.\  D {\bf 74} (2006) 045018
  [arXiv:hep-th/0602026].
  %%CITATION = PHRVA,D74,045018;%%
\bibitem{gilest}
R. Giles and C. B. Thorn, {\sl Phys. Rev.} {\bf D16} (1977) 366.
\bibitem{casher}
A.~Casher, Phys.\ Rev.\ D {\bf 14} (1976) 452.
\bibitem{thornfishnets}
  C.~B.~Thorn,
  %``On The Derivation Of Dual Models From Field Theory.  2,''
 Phys.\ Lett.\ B {\bf 70} (1977) 85, Phys.\ Rev.\ D {\bf 17} (1978) 1073.
\bibitem{thornfreedom}
C.~B.~Thorn,
  %``Asymptotic Freedom In The Infinite Momentum Frame,''
  Phys.\ Rev.\ D {\bf 20} (1979) 1934.
  %%CITATION = PHRVA,D20,1934;%%
\end{thebibliography}
\end{document}